\documentclass[12pt]{iopart}
\usepackage{amssymb}
\usepackage{amsopn}
\usepackage{graphics}
\usepackage{graphicx}
\usepackage{bm}
\newcommand{\eee}{\makebox{e}}
\newcommand{\sign}{\makebox{sign}}

\newcommand{\dd}[0]{\mathrm{d}}
\newcommand{\bb}[0]{\begin{eqnarray}}
\newcommand{\ee}[0]{\end{eqnarray}}
\newcommand{\nn}{\nonumber}
\newcommand{\fd}{\frac{1}{2}}

\newcommand{\prodd}[2]{\overrightarrow{\prod_{#1}^{#2}}}
\newcommand{\prodg}[2]{\overleftarrow{\prod_{#1}^{#2}}}

\newcommand{\action}{\mathcal{S}}
\newcommand{\Sising}{\action_{\mathrm{Ising}}}
\newcommand{\Sint}{\action_{\mathrm{int}}}

\newcommand{\vk}{{\bm{k}}}
\newcommand{\vp}{{\bm{r}}}
\newcommand{\bet}{\alpha'}
\DeclareMathOperator*{\TR}{\mathfrak{Tr}}
\DeclareMathOperator*{\sTR}{\mathsf{Tr}}

\begin{document}
\title[]{Alternative description of the 2D Blume-Capel model using
Grassmann algebra}
\author{Maxime Clusel$^{\dag} $, Jean-Yves Fortin$^\ddag$ and Vladimir N.
Plechko$^\P$}
\address{$^\dag$\ Department of Physics and Center for Soft Matter
Research, New-York University, 4 Washington place, New-York NY 10003, USA}
\address{$^\ddag$\
Laboratoire de Physique Th\'eorique, CNRS/UMR 7085 and Universit\'e
Louis Pasteur, 3 rue de l'Universit\'e, 67084 Strasbourg cedex, France}
\address{$^\P$\ Bogoliubov Laboratory of Theoretical Physics, Joint
Institute for Nuclear Research, 141980 Dubna,
Moscow Region, Russia}
\eads{\mailto{mc2972@nyu.edu}, \mailto{fortin@lpt1.u-strasbg.fr},
\mailto{plechko@thsun1.jinr.ru}}

\begin{abstract}
We use Grassmann algebra to study the phase transition in the
two-dimensional ferromagnetic Blume-Capel model from a fermionic point of
view. This model presents a phase diagram with a second order critical line
which becomes first order through a tricritical point, and was used to
model the phase transition in specific magnetic materials and liquid
mixtures of He$^3$-He$^4$. In particular, we are able to map the spin-1
system of the BC model onto an effective fermionic action from which we
obtain the exact mass of the theory, the condition of vanishing mass
defines the critical line. This effective action is actually an
extension of the free fermion Ising action with an additional quartic
interaction term. The effect of this term is merely to render the
excitation spectrum of the fermions unstable at the tricritical point.
The results are compared with recent numerical Monte-Carlo simulations.
\end{abstract}
\submitto{\JPA}


%
%
%
\section{Introduction}

The Blume-Capel (BC) model is a classical spin-1 model, originally
introduced to study phase transitions in specific magnetic materials
with a possible admixture of non-magnetic states \cite{blume66,capel66}.
Its modification  was also used to qualitatively explain the phase
transition in a mixture of He$^3$-He$^4$ adsorbed on a two-dimensional
(2D) surface \cite{BeG71}.  Below a concentration of 67\% in He$^3$, the
mixture undergoes a so-called $\lambda$ transition:  the two components
separate through a first order phase transition and only He$^4$ is
superfluid. On a 2D lattice representing an Helium film, He atoms are
modelled by a spin-like variable, according to the following rule: an
He$^3$ atom is associated to the value 0, whereas a He$^4$ is represented
by a classical Ising spin taking the values $\pm 1$. Within this framework,
all the lattice sites are occupied either by an He$^3$ or He$^4$ atom
\cite{BeG71}. The 2D Blume-Capel model describes the
behaviour of this ensemble of spins $\{S_{mn}^{}=0,\pm1\}$. In addition to
the usual nearest-neighbour interaction, its energy includes the term
$\Delta_0 \sum_{mn}S_{mn}^2$, with $S_{mn}^{2}=0,1$, to take into account
a possible change in vacancies number. $\Delta_0$ can be thought as a
chemical potential for vacancies, or as a parameter of crystal field in a
magnetic interpretation.  A simple analysis of the 2D BC Hamiltonian
already shows that this model presents a rather complex phase diagram in
the plane $(T,\Delta_0)$, where $T$ is the temperature in the canonical
ensemble \cite{cardy96}. In the limit $\Delta_0\rightarrow -\infty$, the
values $S_{mn}=0$ are effectively excluded and the standard 2D Ising model
is recovered, with its well-known second-order critical point at
$(T,\Delta_0)=(T_\mathsf{c}=2/\ln(1+\sqrt{2})\simeq
2.\,269185,-\infty)$, with the parameters taken in units of the Ising
exchange energy $J$. At zero temperature, on the other hand, a simple
energy argument shows that the ground state is the Ising like ordered state
with $|S_{mn}|=1$ if $\Delta_0<2$, and $|S_{mn}|=0$ else. There is
therefore a first order phase transition at $(T,\Delta_0)=(0,2)$,
suggesting a change in the order of the transition at some tricritical
point at the critical line at finite temperatures. Mean field theory
confirms this behaviour, and provides a second order transition line in the
plane $(T,\Delta_0)$ in the region extending from negative to moderated
positive values of $\Delta_0$ \cite{blume66,capel66,BeG71, cardy96,
hoston91}.  Beyond the tricritical point, as dilution increases, the
transition becomes first order. Precise numerical simulations have been
performed to study the phase diagram and to locate the tricritical point
of the 2D Blume-Capel model \cite{beale86, xalap98, liusch02, dasilva02,
silva06}. From a theoretical background, several approximations have been
used as well, such as mean field theory \cite{blume66, capel66, BeG71,
cardy96, hoston91}, renormalization group analysis \cite{burk76,berk76},
and high temperature expansions \cite{saul74}. Using correlation identities
and Griffith's and Newman's inequalities, rigorous upper bounds for the
critical temperature have been obtained by Braga \textit{et al}
\cite{braga94}. It was also conjectured that exactly at the tricritical
point the 2D BC model falls into the conformal field theory (CFT) scheme of
classification of the critical theories in two dimensions
\cite{belavin84,friedan84,senechal99}. This is the case with $m=4$ and
$c=7/10$, where $c$ is the central charge \cite{friedan84, senechal99,
chenkel93}. The CFT analysis also implies a specific symmetry called
supersymmetry in the 2D BC model at the tricritical point \cite{friedan84,
senechal99,chenkel93}.\\
The two-dimensional BC model is directly related
as well to percolation theory \cite{denguo05} and dilute Potts model
\cite{qideng05}, where tricritical point properties are observed for
percolating clusters of vacancies. We also mention quantitative results
that match the universality class at the tricritical point of the BC model
with the one of a 2D spin fluid model representing a magnetic gas-fluid
coexistence transition \cite{wilding96}, and similarities between BC phase
diagram and Monte-Carlo results on the extended Hubbard model on a square
lattice \cite{paw06}.  The advanced theoretical methods like bootstrap
approach and perturbed conformal analysis, in combination with the
integrable quantum field theory and numerical methods, have been applied to
study the scaling region and the RG flows in the 2D BC universality class
\cite{demusi95,fimusi20, fimusi201}.

The aim of this article is to present a different analytical method for the
BC model in two dimensions with the use of the anticommuting Grassmann
variables, originally proposed for the classical 2D Ising model in the case
of free fermions \cite{ple85dok, ple85tmp} and since then used to treat
various problems around the 2D Ising model, such as finite size effects and
boundary conditions \cite{liaw99,wuhu02}, quenched disorder
\cite{ple98,ple98fl}, and boundary magnetic field \cite{clusel05,clusel06}.
In contrast with the use of more traditional combinatorial and
transfer-matrix considerations \cite{ber69, samuel80, itz82, nojima98,
dritz89}, this method is rather based on a direct introduction of Grassmann
variables (fermionic fields) into the partition function $Z$ in order to
decouple the spin degrees of freedom in the local bond Boltzmann weights in
$Z$. A purely fermionic integral for $Z$ then follows by eliminating spin
variables in the resulting mixed spin-fermion representation for $Z$. The
method turns out to be particularly efficient to deal with models with
nearest-neighbour interactions in the 2D plane \cite{ple85dok,ple85tmp}.
For the 2D Ising model, the fermionic integral for $Z$ appears to be a
Gaussian integral over Grassmann variables, with the quadratic fermionic
form (typically called action) in the exponential \cite{ple98,dritz89}.
Respectively, the model is exactly solvable by Fourier transformation to
the momentum space for Grassmann variables in the action. In physical
language, this corresponds to the case of free fermions
\cite{dritz89,ple95amm}.

As the additional crystal field term in the BC Hamiltonian is local, we
hoped that the method will be applicable as well in this context. We will
see in the following that though it is not possible to compute exactly the
partition function and thermodynamics quantities of the BC model directly,
since the resulting fermionic action for BC is non Gaussian, our approach
allows to derive in a controlled way physical consequences from the
underlying fermionic lattice field theory with interaction. In the
continuum limit, a simplified effective quantum field theory can be
constructed and analyzed in the low energy sector, leading to the exact
equation for the critical line, that follows from the condition of
vanishing mass, and to the effective interaction between fermions
responsible for the existence of a tricritical point. The effects of
interaction are assumed to be analyzed in the momentum-space
representation. An approximate scheme such as Hartree-Fock-Bogoliubov (HFB)
method can be used to locate the tricritical point. There are also some
albeit formal analogies in this respect with the approaches typically used
in the BCS theory of ordinary superconductivity.  In general, it is
interesting to note that in 2D a phase diagram of the BC model with first
order transition and tricritical point can be described not only within a
bosonic $\Phi^6$ Ginzburg-Landau theory \cite{lawrie84,zj04}, where the
order parameter is a simple scalar, but also with the use of fermionic
variables.

The article is organized as follows. After presenting the BC Hamiltonian
and the related partition function in standard spin-1 interpretation, we
apply the fermionization procedure leading to the exact fermionic action on
the lattice. Then, from this result, we derive the effective action in
the continuum limit and extract the exact mass. The condition of zero mass
already gives the equation for the critical line in the $(T,\Delta_0)$
plane. The effective action also includes four-fermion interaction due to
admixture of the $S^{2} =0$ states (vacancies) in the system, with coupling
constant $g_0\propto \exp(-\Delta)$, where $-\Delta =\Delta_0/T$, and
$\Delta_0$ is the parameter of the crystal field in the Hamiltonian. We
then give a physical interpretation for the existence of a tricritical
point in  the BC phase diagram by studying the fermionic stability of the
BC spectrum at the critical line at order ${\bf k}^2$ in momentum and
compare our results with recent numerical Monte Carlo simulations
\cite{beale86,xalap98,liusch02,dasilva02,silva06}.

\section{The 2D Blume-Capel model}
\subsection{Hamiltonian and partition function}

The 2D BC model is defined, on a square
lattice of linear size $L$, via the following Hamiltonian:
\bb
\fl
H = -\sum_{m=1}^{L}\sum_{n=1}^{L}\Big[J_1 S_{mn}S_{m+1n}
+J_2 S_{mn}S_{mn+1}\Big] +\Delta_0\sum_{m=1}^{L}
\sum_{n=1}^{L}S_{mn}^{2}\, . \;\;\;
\label{ham1a}
\ee
In the above expression, $S_{mn}=0,\pm1$ is the BC spin-1 variable
associated with the $mn$ lattice site, with $m,n=1,2,3,\ldots,L$, where
$m,n$ are running in the horizontal and vertical directions, respectively.
The total number of sites and spins on the lattice is $L^2$, at final
stages $L^2\to\infty$. The spins are interacting  along the lattice bonds,
$J_{1,2}>0$ are the exchange energies. Notice that positive $J_{1,2}>0$
correspond to the ferromagnetic case. In addition to the Ising states with
$S_{mn}=\pm1$, there are as well the non-magnetic atomic levels with
$S_{mn}=0$, which we shall also refer to as vacancies.  The crystal field
parameter $\Delta_0$ plays the role of a chemical potential, being
responsible for the level splitting between states $S_{mn}=0$ and
$S_{mn}=\pm 1$. The Hamiltonian that appears in the Gibbs exponential
may be written in the form:

\bb
\fl
-\beta H =\sum_{m=1}^{L}\sum_{n=1}^{L}\Big[K_1 S_{mn}S_{m+1n}
+K_2 S_{mn}S_{mn+1}\Big] +\Delta\sum_{m=1}^{L}
\sum_{n=1}^{L}S_{mn}^{2}\, ,\;\;\;
\label{ham1ab}
\ee
where $K_{1,2} =\beta J_{1,2}$ are now the temperature dependent coupling
parameters, $\beta =1/T$ is inverse temperature in the energy units,
and $\Delta=-\beta\Delta_0$. In what follows we will assume, in
general, the ferromagnetic case, with positive $J_{1,2}>0$ and $K_{1,2}>0$,
though the fermionization procedure by itself is valid irrespective of
the signs of interactions. \footnote{ \
In what  follows, by presenting the numerical results, we shall typically
assume isotropic case for interactions in the above Hamiltonians, with
$J_1=J_2=J,\; K_1=K_2=K$, and $K=\beta J$. We will also use, in some
cases, the dimensionless parameters normalized by the exchange energy $J$
for temperature $T$ and chemical potential $\Delta_0$.}\, The positive
$\Delta$ (negative $\Delta_0$) is favourable for the appearance of the
Ising states with $S_{mn}^{2}=1$ in the system, with the ordered phase
below the critical line in the $(T,\Delta_0)$ plane, at low temperatures,
while negative $\Delta$ (positive $\Delta_0$) will suppress Ising states,
being favourable for vacancies. In the limit $\Delta\to\infty$, or
$\Delta_0\to-\infty$, the states with $S_{mn}^{2}=0$ are effectively
suppressed and the model reduces to the 2D Ising model, with the critical
temperature being defined by the condition $\sinh 2K_1\sinh 2K_2 =1$.  As
$\Delta_0$ increases to finite values, there will be a line of phase
transitions in the $(T,\Delta_0)$ plane. The increasing $\Delta_0$ admits
the appearance of the vacancy $S_{mn}^{2}=0$ states. Respectively, the
critical line goes lower as $\Delta_0$ increases from negative to positive
values and terminates at $\Delta_0=J_1+J_2$ at zero temperature, so that
all sites are empty at larger positive values of $\Delta_0$ at $T=0$. A
remarkable feature of the BC model is that there is also a tricritical
point on the critical line at finite temperatures somewhere slightly to the
left from $\Delta_0=J_1+J_2$, where the transition changes from second to
first order.\\
The partition function $Z$ of the BC model is obtained by summing over all
possible spin configurations provided by $\{S_{mn}=0,\pm1\}$ at each site,
$Z= \sum_{S=0,\pm1} e^{-\beta H} =\sTR_{\{ S \}} e^{-\beta
H}$. Using the property $\{S_{mn}=0,\pm1\}$, it is easy to develop each
Boltzmann factor appearing in the above trace formula in a polynomial form:
\bb
\exp \left( K_{\,i} S S'\right)
=1+\lambda_{\,i} S S'
+\lambda_{\,i}' S^2 S'^2, \;\;\;\;
\,i=1,2\,,\;\;
\label{defpol}
\ee
with
\bb
\lambda_{\,i}=\sinh K_{\,i}\,,\;\;\;\;
\lambda_{\,i}'=\cosh K_{\,i} -1\,. \;\;\;\; \,i=1,2\,.\;\;
\label{deflambda}
\ee
The partition function is then given by the product of the above
spin-polynomial Boltzmann weights under the averaging:

\bb
\fl
Z=\sTR_{\{S_{mn}=0,\pm1\}}
\Big\{\prod_{m=1}^{L}\prod_{n=1}^{L} e^{\,\Delta S_{mn}^{2}}\,
\Big[
(1+ \lambda_1\, S_{mn}S_{m+1n}
+\lambda_{1}'S_{mn}^{2}S_{m+1n}^{2}) \nn \\
\times\,(1+ \lambda_2\, S_{mn}S_{mn+1}
+\lambda_{2}'\, S_{mn}^{2}S_{mn+1}^{2}) \Big]\Big\}.\;\;
\label{PFspin}
\ee
This expression will be the starting point of the fermionization procedure
for $Z$ using Grassmann variables we develop in Section 3. At first stage,
we introduce new Grassmann variables to decouple the spins in the local
polynomial factors of expression (\ref{PFspin}). At next stage, we
sum over spin states in the resulting mixed spin-fermion representation for
$Z$ to obtain a purely fermionic theory for $Z$.

\subsection{Local spin decomposition}

In what follows, we shall need to average partially fermionized $Z$ over
the spin states at each site.
This averaging will be performed in two steps, first we keep in mind to
average over the Ising degrees of freedom, $S_{mn}=\pm1$, then adding the
contribution of vacancies, $S_{mn}=0$. The two cases may be also
distinguished in terms of variable $S_{mn}^{2}=0,1$. In this subsection, we
shortly comment on the formalization of this two-step averaging. Provided
we have any function of the BC spin-1 variable $f(S_{mn})$, with
$S_{mn}=0,\pm1$, the averaging rule is simple:
\bb
\sum_{S_{mn}=0,\pm1} \,f(S_{mn}) =f(0)+f(+1)+f(-1)\,.\;\;
\label{fuy1}
\ee
In forthcoming procedures, we ought to average first over the states
$S_{mn}=\pm1$ at each site, provided $S_{mn}^{2}=1$, while making the sum
over choices $S_{mn}^{2}=0,1$ at next stage. In principle, since $S_{mn}
=\sign\{ S_{mn}\}\,|S_{mn}|$, with $\sign\{S_{mn}\}=\pm1$ and $|S_{mn}|
=S_{mn}^{2} =0,1$, we can try simply to write $S_{mn} =y_{mn}\sigma_{mn}$,
where $y_{mn}=0,1$, and $\sigma_{mn}=\pm1$, and to average over the
component states $y_{mn}=0,1$ and $\sigma_{mn} =\pm1$ as independent
variables. This gives:
\bb\fl
\sum_{y_{mn}=0,1;\, \sigma_{mn}=\pm1} \,f(y_{mn}\sigma_{mn})
=f(+0) +f(-0)+f(+1)+f(-1)\,.\;\;
\label{fuy2}
\ee
We see that the zero state is counted twice, in contradiction to
(\ref{fuy1}). This may be corrected by introducing in the definition of
the averaging the weight factor $\frac{1}{2}$ at $y_{mn}=0$. Equivalently,
this may be formalized by adding $2^{\,-1+y_{mn}}$ under the sum. This
results the sum of three terms in agreement with (\ref{fuy1}):
\bb\fl
\sum_{y_{mn}=0,1;\,\sigma_{mn}=\pm1}\,2^{-1+y_{mn}}\,
f(\sigma_{mn}y_{mn}) =f(0)+f(+1)+f(-1)\,.\;\;
\label{fuy3}
\ee
In fact, this decomposition scheme with $S_{mn} =\sigma_{mn}\,y_{mn}$ and
independently varying $\sigma_{mn}=\pm1$  and $y_{mn}=0,1$ is somewhat
more close to the situation for the two-dimensional Ising model with
quenched site dilution \cite{ple98,ple98fl}. In that case
$\sigma_{mn}=\pm1$ is simply the Ising spin, while the variable
$y_{mn}=0,1$ is the quenched dilution parameter, counting whether the given
site is  occupied or dilute, and both averaging rules (\ref{fuy2}) and
(\ref{fuy3}) can be interpreted physically. The case (\ref{fuy2}) means
in fact that there is a spin $\sigma_{mn}=\pm 1$ also at site $y_{mn}=0$,
which is not interacting with its nearest-neighbors. This
empty, or rather disconnected site, by flipping over two states $\pm1$
under temperature fluctuations, will give however a contribution to the
entropy, $\ln 2$ by empty site. The case (\ref{fuy3}) means that the site
$y_{mn}=0$ is really dilute, or empty, with no spin degree of freedom at
it, even disconnected.
For the quenched dilute 2D Ising model, the
quenched averaging over some fixed temperature-independent distribution
$y_{mn} =0,1$ is physically distinct from the $\sigma_{mn}=\pm1$ averaging,
and is assumed to be performed rather on $-\beta F =\ln Z$, but not on $Z$
itself. The situation is different for the BC model, which is in essence
the annealed case of the site dilute Ising model, with the averaging
simultaneously over all states $S_{mn}^{}=0,\pm1$ at each site for $Z$
itself. In this case the averaging is to be performed strictly according
to the rules like (\ref{fuy1}) and (\ref{fuy3}), but not (\ref{fuy2}).

There is still another way to formalize the averaging over the
possibilities of $S_{mn}=\pm1$ before we actually fix $S_{mn}^{2}=0,1$.
It is based on the observation that the result of the averaging
(\ref{fuy1}) will not be changed if we replace $S_{mn}\to \sigma_{mn}
S_{mn}$, with $\sigma_{mn}=\pm1$, since the sum includes $S_{mn} =\pm1$
anyhow:
\bb
\fl
\sum_{S_{mn}=0,\pm1}\,f(S_{mn})
=\sum_{S_{mn}=0,\pm1} \,f(\sigma_{mn}S_{mn})
=f(0)+f(+1)+f(-1)\,,\;\; \sigma_{mn}=\pm1\,.
\label{fuy4}
\ee
Thought the above equation holds already for any fixed value of
$\sigma_{mn}=\pm1$, we can as well average it over the states
$\sigma_{mn}=\pm1$, introducing factor $\frac{1}{2}$ for normalization.
The averaging of $f(\sigma_{mn}S_{mn})$ itself gives:
\bb
\fl
\frac{1}{2}\sum\limits_{\sigma_{mn}=\pm1}^{} f(\sigma_{mn}S_{mn})
=\frac{1}{2}[f(S_{mn})+f(-S_{mn})] =g(S_{mn}^{2})\,,\;\;\;\;
S_{mn}^{2}=0,1\,.\;\;
\label{fuy5}
\ee
The result of the averaging will be a function $g$ which only depends on
$|S_{mn}|=0,1$, alias $S_{mn}^{2}=0,1$, but not on the $\sign\{S_{mn}\}$
of $S_{mn} =\sign\{S_{mn}\}|S_{mn}|$. In terms of $g(S_{mn}^{2})$, the
equation (\ref{fuy4}) results:
\bb
\fl
\sum_{S_{mn}=0,\pm1}\,f(S_{mn})
=\sum_{S_{mn}=0,\pm1}\Big\{
\frac{1}{2}\sum_{\sigma_{mn}=\pm1}
f(\sigma_{mn}S_{mn}) \Big\}
=g(0) +2g(1)\,.\;\;
\label{fuy6}
\ee
In this form the two-step averaging will be realized in the procedure
of elimination of spin variables by constructing the fermionic integral
for $Z$ in the forthcoming discussion.

%
%
\section{Fermionization and lattice fermionic field theory}

The expression of the BC partition function $Z$ as a product of spin
polynomials under the averaging as given in (\ref{PFspin}) will be the
starting point of the fermionization procedure for $Z$. This procedure
has first been introduced in the context of the 2D pure Ising model
\cite{ple85dok, ple85tmp}. It relies on interpreting each spin polynomial
Boltzmann weight in (\ref{PFspin}) as the result of integration over a set
of two Grassmann variables, which decouples the spins under the integral.
Before going into details, we remind in the following subsection few
essential features about Grassmann variables and the rules of integration.

\subsection{Grassmann variables}

Mathematically, Grassmann variables may be viewed as formal purely
anticommuting fermionic numbers \cite{ber66}. In physical aspect, they are
images of quantum fermions in path integral \cite{ber66}. We remind here
few basic features about Grassmann variables that are needed in the
rest of the paper. More details can be found in \cite{nakahara99,zinn05}.
A Grassmann algebra $\mathcal{A}$ of size $N$ is generated by a set of $N$
anti-commuting objects $\{a_1,a_2,\ldots, a_N$\} satisfying:
\bb
a_ia_j +a_ja_i =0\,,\;\;\; a_{i}^{2}=0\,,\;\;\;\;
i,j =1,2,\ldots,N\,.\;\;
\label{grass1}
\ee
This as well implies $a_ia_j =-a_ja_i$, including the case $i=j$.
Unlike quantum fermions, Grassmann variables are totally anticommuting.
Note that any linear superpositions of the original variables
(\ref{grass1}) are again purely anti-commuting with each other and with the
original variables, and their squares are zeroes. Functions defined on such
an algebra are particularly simple, they are always polynomials with a
finite degree (since $a_{i}^{2}=0$).  It is possible to define the notion
of integration \cite{zj04, ber66,nakahara99,zinn05} in algebra of such
polynomials with the following rules. For one variable, the rules are:
\bb
\int \dd a_i\cdot a_i=1\,,\;\;\;\;\;
\int \dd a_i\cdot 1=0\,.\;\;
\ee
The integral with many variables is considered as a multilinear functional
with respect to each of the variables involved into integration
(integral of a sum is the sum of the integrals etc). In multiple integrals,
the fermionic differentials are assumed again anti-commuting with each
other and with the variables themselves. The integration of any polynomial
function of Grassmann variables like $f(a) =f(a_1,a_2,\ldots,a_N)$ then
reduces, in principle, to a repeating use of the above rules.
The rules  of change of variables in Grassmann variable (fermionic)
integrals under a linear substitution are similar to the analogous rules of
common (commuting) analysis. The only difference is that the Jacobian of
the transformation will enter now in the inverse power, as compared to the
commuting (bosonic) case \cite{zj04, ber66, nakahara99, zinn05}.\\
With the above definitions, the Gaussian integrals over Grassmann variables
are all expressed by the equations relating them to determinants and
Pfaffians. The basic equation for the determinantal integral of first kind
reads:
\bb
\int \prod_{i=1}^{N}\dd a_i^* \dd a_i \exp\left(\sum_{i,j=1}^{N}
a_iA_{ij} a^*_j\right)=\det A\,,\;\;
\label{deta1}
\ee
where the integration is over the doubled set of totally anti-commuting
variables $\{a,\,a^*\}$. The (square) matrix $A$ in the exponential is
arbitrary. In applications, the quadratic fermionic form in the exponential
like in (\ref{deta1}) is typically called {\em action}. Since the action is
quadratic, the integral is Gaussian. The exponential in (\ref{deta1}) is
assumed in the sense of its series expansion. Due to nilpotent properties of
fermions, the exponential series definitely terminates at some stage,
thus resulting a finite polynomial in variables involved under the
integral. [With respect to the action $S=aAa^*$ taken as a whole, the last
nonzero term will be with $S^N\neq  0$, while $S^{N+1}=0$.  Alternatively,
the same polynomial for the exponential from (\ref{deta1}) will follow by
multiplying elementary factors like $\exp(a_iA_{ij}a_{j}^{*})=
1+a_iA_{ij}a_{j}^{*}$]. In physical interpretation, the integral of the
first kind (\ref{deta1}) with complex-conjugate fields rather corresponds
to Dirac theories.\\
The Majorana theories with real fermionic fields are presented by the
Gaussian integrals of the second kind related to the Pfaffian. The basic
identity for the fermionic integral of the second kind reads:
\bb
\int \prodg{i=1}{N} \dd a_i \exp\left(\sum_{i,j=1}^{N}
\frac{1}{2}\,a_iA_{ij}a_j\right)
=\mbox{Pf}\,A\,.\;\;
\label{pfaff1}
\ee
The integration is over the set of even number $N$ of Grassmann
variables, the arrow in the measure indicates the direction of ordering
of anti-commuting differentials. The matrix in the exponential is now
assumed skew-symmetric, $A_{ij}+A_{ji}=0,\;\; A_{ii}=0$, which
property is complimentary to fermionic anticommutativity. The result of
the integration is the Pfaffian associated with the
skew-symmetric matrix $A$ from the exponential, otherwise, one can
associate the Pfaffian on the r. h. side of (\ref{pfaff1}) with the
above-diagonal triangular array of elements of that matrix,
$\{A_{ij}\,|\, 1\leq i< j \leq N\}$. In mathematics, the Pfaffian is known
as a certain skew-symmetric polynomial in elements of a triangular array
of the above kind.  In physics, the  combinatorics of the Pfaffian
also is known under the name of the (fermionic) Wick's theorem.
Note that the identity (\ref{pfaff1}) can be assumed by itself for
the definition of the Pfaffian.\\
In a combinatorial sense, the determinant is rather a particular case of
the Pfaffian. Respectively, the integral (\ref{deta1}) is a subcase of the
integral (\ref{pfaff1}). It can be shown, on the other hand, that
$(\mbox{Pf} \, A)^2 =\det A$ for any skew-symmetric matrix $A$. This
implies that, in principle, an integral of the second kind (\ref{pfaff1})
can always be reduced to an integral of first kind (\ref{deta1}) by
doubling the number of fermions in (\ref{pfaff1}). In applications like in
the Ising and BC models, where the original integrals in the real lattice
space rather appear in the Pfaffian like form of (\ref{pfaff1}), this
reduction to the determinantal case occurs automatically after the
transformation to the momentum space, where the fermionic variables are
typically combined into groups of variables with opposite momenta $(\bm{k},
-\bm{k})$, which play the role of the conjugated variables like in
(\ref{deta1}). In practice, for low-dimensional integrals, most of
calculations can be performed simply from the definition of the integral,
by expanding the integrand functions into polynomials.

\subsection{Fermionization procedures}
In the same spirit as for the 2D Ising model \cite{ple85tmp}, we introduce
two pairs of Grassmann variables $(a_{mn}, \bar{a}_{mn})$ and
$(b_{mn}, \bar{b}_{mn})$ to factorize the polynomials appearing in
(\ref{PFspin}). Namely we use the relations:

\bb
\fl \nn
1+\lambda_1S_{mn}S_{m+1n} +\lambda_1'
S_{mn}^{2}S_{m+1n}^{2}=\int \dd \bar{a}_{mn} \dd a_{mn} \,
e^{(1+\lambda_1'S_{mn}^{2}S_{m+1n}^{2})\, a_{mn}^{}\bar{a}_{mn}}
\\ \nn
\times
(1+a_{mn}S_{mn})\,(1+\lambda_{1}\,\bar{a}_{mn}S_{m+1n}),
\ee
\bb
\fl \nn
1+\lambda_2S_{mn}S_{mn+1} +\lambda_2'
S_{mn}^{2}S_{mn+1}^{2}=\int \dd \bar{b}_{mn} \dd b_{mn}\,
e^{(1+\lambda_2'S_{mn}^{2}S_{mn+1}^{2})\, b_{mn}^{}\bar{b}_{mn}}
\\
\times
(1+b_{mn}S_{mn})\,(1+\lambda_{2}\,\bar{b}_{mn}S_{mn+1}).
\label{fact1}
\ee
For the sake of simplicity in notation we introduce the following
link factors:

\bb\nn
& A_{mn}=1+a_{mn} S_{mn}\,,\;\;
& \bar{A}_{m+1n}=1+\lambda_1 \bar{a}_{mn}
S_{m+1n}\,,\;\;\\
& B_{mn}=1+b_{mn} S_{mn}\,,\;\;
& \bar{B}_{mn+1}=1+\lambda_2 \bar{b}_{mn}S_{mn+1}\,. \;\;
\label{link1}
\ee
We also define the Grassmann local trace operators which associate to any
function $f(\ldots)$ on the Grassmann algebra as follows:
\bb
\nn
\TR_{(a_{mn})} \big[ f(a_{mn},\bar{a}_{mn})\big]
=\int \dd \bar{a}_{mn} \dd a_{mn} \,
e^{(1+\lambda_1'S^2_{mn}S^2_{m+1n})\, a_{mn}^{}\bar{a}_{mn}^{}}
f(a_{mn},\bar{a}_{mn}), \\
\TR_{(b_{mn})} \big[ f(b_{mn},\bar{b}_{mn}) \big]
=\int \dd \bar{b}_{mn} \dd b_{mn}\,
e^{(1+\lambda_2'S^2_{mn}S^2_{mn+1})\,
b_{mn}^{}\bar{b}_{mn}^{}}\, f(b_{mn},\bar{b}_{mn}).
\label{fact2}
\ee
The factorized Boltzmann weights from (\ref{fact1}) now read:

\bb
\nn
1+\lambda_1S_{mn}S_{m+1n}+\lambda_1'
S_{mn}^2S_{m+1n}^2=\TR_{(a_{mn})} \big[ A_{mn}
\bar{A}_{m+1n}\big],\\
1+\lambda_2S^2_{mn}S^2_{mn+1}+\lambda_2'
S^2_{mn}S^2_{mn+1}=\TR_{(b_{mn})} \big[B_{mn} \bar{B}_{mn+1}
\big].
\label{fact3}
\ee
Introducing the above Boltzmann weights into the original expression
(\ref{PFspin}) for $Z$, we obtain a mixed representation containing both
spins and Grassmann variables for $Z$. Notice that as the separable link
factors like $A_{mn}, \bar{A}_{mn}, B_{mn}, \bar{B}_{mn}$ are
neither commuting nor anti-commuting with each other, the order in which
they appear in the product may be important. The factorized bond weights,
however, presented in (\ref{fact3}) by doubled link factors under the trace
operators, are totally commuting, if taken as a whole, with any element of
the algebra under the averaging. For the whole lattice, following the rules
(\ref{fact2}), we define the global trace operator as follows:
\bb
\label{fact4} \fl
\TR_{(a,b)}\, \big[ f \big]
=\int  \prod_{m=1}^L \prod_{n=1}^L \dd \bar{a}_{mn}
\dd a_{mn}\dd \bar {b}_{mn} \dd b_{mn} e^{\Delta S^2_{mn}}
f(a_{mn},\bar{a}_{mn},b_{mn},\bar{b}_{mn})\\ \nn \times \exp \left\{
\sum_{m=1}^L \sum_{n=1}^L \left[ (1+\lambda'_1 S^2_{mn}
S^2_{m+1n}) a_{mn} \bar{a}_{mn} +(1+\lambda'_2 S^2_{mn}
S^2_{mn+1}) b_{mn} \bar{b}_{mn}\right] \right\}.
\ee
The all even-power terms in spin variables are now incorporated
into the generalized Gaussian averaging measure of (\ref{fact4}),
including the term with chemical potential. The partition function is
then given by

\bb
\label{PFmix}
Z=\sTR_{\{S\}}
\TR_{(a,b)} \left[ \prodd{n=1}{L} \left( \prodd{m=1}{L}
\left( (A_{mn}\bar{A}_{m+1n}) (B_{mn} \bar{B}_{mn+1})
\right) \right)
\right].
\ee
At this stage the factorized partition function appears as a double trace,
over the spin degrees of freedom, with $\sTR_{\{S\}}$, and over the
Grassmann variables, with $\TR_{(a,b)}\,$. The idea of the next step is to
make spin summation in (\ref{PFmix}) to obtain a purely fermionic integral
for $Z$. At first stage, we keep in mind to eliminate rather the Ising
degrees $\pm1$ of spin variables in (\ref{PFmix}). The averaging over
$S_{mn}^{2}=0,1$ will be performed at next stage.

\subsection{The ordering of factors}

Up to now we only add extra fermionic (Grassmann) variables to obtain the
mixed expression (\ref{PFmix}), where the spin variables are actually
decoupled into separable link factors like (\ref{link1}). Further algebraic
manipulations are necessary to simplify this expression in order the spin
averaging be possible in each group of factors with the same spin.  For any
given $mn$, there are four such factors, $A_{mn}, B_{mn}, \bar{A}_{mn},
\bar{B}_{mn}$, which all include the same BC spin $S_{mn}=0, \pm1$. What we
need is to be able to keep nearby the above four factors with the same spin
at least at the moment of the spin averaging.  We apply the mirror-ordering
procedure, introduced originally for the two-dimensional Ising model, to
move together, whenever possible, the different link factors containing the
same spin. Despite of that the separable link factors like (\ref{link1})
are in general neither commuting nor anticommuting, it is still possible to
make use of the property that the doubled combinations like $A_{mn}
\bar{A}_{m+1n}$ and $B_{mn} \bar{B}_{mn+1}$ are effectively commuting, if
taken as a whole, with any element of the algebra under the sign of the
Gaussian fermionic averaging in (\ref{PFmix}).  Using the notation for the
ordered products similar to that of \cite{ple85dok,ple85tmp}, this leads
to:
\bb
\nn
Z&=& \sTR_{\{S\}} \TR_{(a,b)}\Big\{\prod\limits_{m=1}^{L}
\prod\limits_{n=1}^{L}
\Big[ (A_{mn}\bar{A}_{m+1n})(B_{mn}\bar{B}_{mn+1})
\Big]\Big\}
\\ \nn
&=& \sTR_{\{S\}} \TR_{(a,b)}
\Big\{ \prodd{n=1}{L}\, \Big[\, \prodd{m=1}{L}
\bar{B}_{mn}A_{mn}\bar{A}_{m+1n}\cdot \prodg{m=1}{L}B_{mn}\, \Big] \Big\}
\;\;\;\;
\\
\label{PFmixmo}
&=&
\sTR_{\{S\}} \TR_{(a,b)} \left[
\prodd{n=1}{L} \left( \left( \prodd{m=1}{L}
\bar{A}_{mn} \bar{B}_{mn} A_{mn} \right)\cdot \left(\prodg{m=1}{L} B_{mn}
\right) \right) \right].
\ee
In the above transformations, we use mirror-ordering decoupling for factors
in vertical direction, $B_{mn}\bar{B}_{mn+1}$, with respect to $n$, then
insert the commuting factorized horizontal weights, $A_{mn}\bar{A}_{m+1n}$,
and reread the resulting products in few subsequent transformations
(cf. \cite{ple85dok,ple85tmp}). The boundary terms are also to be treated
properly as we pass from (\ref{PFspin}) to (\ref{PFmixmo}). The simplest
case is provided by the free boundary conditions.  The free boundary
conditions for spin variables, $S_{L+1n} =S_{mL+1}=0$, in (\ref{PFspin})
correspond to the free boundary conditions for fermions, $\bar{a}_{0n}
=\bar{b}_{m0} =0$,  in (\ref{PFmixmo}). For free boundary conditions, the
transformation from (\ref{PFspin}) to (\ref{PFmixmo}) is exact. In what
follows, however, we will typically assume the periodic boundary conditions
for fermions in representations like (\ref{PFmixmo}). These are most
suitable closing conditions when passing to the Fourier space for
anticommuting (Grassmann) fields. The change of the boundary conditions of
this kind is inessential in the limit of infinite lattice as $L^2\to
\infty$. In principle, one can pay more attention to the effects of the
boundary terms in the periodic case, which can actually be treated
rigorously also for finite lattices \cite{ple85tmp, liaw99, wuhu02,
clusel05, clusel06}.\\
In the case of the 2D Ising
model, with $S_{mn}^{2}=1$, we can explicitly perform the trace over the
Ising spin degrees of freedom $S_{mn}=\pm1$ recursively at the junction of
two $m$-ordered products in the final line of (\ref{PFmixmo}). The
situation is slightly different in the BC case, since $S_{mn}^{2}=0,1$,
instead.  Also, the trace operator (\ref{fact4}) contains terms with
$S_{mn}^2=0,1$ which are coupled at neighboring sites. Therefore it is not
possible to trace over the whole set of states $S_{mn} =0,\pm1$ in the BC
case directly in (\ref{PFmixmo}), but only we can eliminate first the Ising
degrees $\sign\{S_{mn}\}=\pm1$. The BC variables $S_{mn}^{2}=0,1$ will
still remain as parameters and will be eliminated at next stages. The
elimination of the Ising degrees we will realize by the symmetrization
transformation, $S_{mn}\to \sigma_{mn}S_{mn}$, with averaging over
$\sigma_{mn} =\pm1$, following the procedures explained in
(\ref{fuy4})-(\ref{fuy6}) above. The details of the $\sigma_{mn}=\pm1$
averaging are discussed in the next subsection. Thus, the ordering
procedure on the link variables (\ref{link1}) allows us to eliminate
at least a one part of the spin degrees of freedom in the factorized
expression for $Z$ resulting in the final line of (\ref{PFmixmo}).

\subsection{Spin summation}

At the junction of the two ordered products in (\ref{PFmixmo}), with
$S_{mn}\to \sigma_{mn}S_{mn}$, we perform the trace $\sigma_{mn} =\pm1$
recursively, for $m=L,L{-}1,\ldots,2,1$, for given fixed $n$, starting with
$m=L$. The procedure will be then repeated for other values of
$n=1,2,3,\ldots,L$.  The four relevant factors $\bar{A}_{mn},\bar{B}_{mn},
A_{mn},B_{mn}$ with the same spin that met at the junction of the two
$m$-product in (\ref{PFmixmo}), for given $n$, are to be specified from
(\ref{link1}). There we assume $S_{mn}\to \sigma_{mn}S_{mn}$.  Then we
multiply the above four factors, taking into account that
$\sigma_{mn}^{2}=1$, so that $S_{mn}^{2} \to \sigma_{mn}^{2}S_{mn}^{2} \to
S_{mn}^{2}$, and sum over the states $\sigma_{mn}=\pm1$. This will
eliminate all odd terms in the polynomial so obtained. The averaging thus
results:
\bb
\fl \nn
\fd\sum_{\sigma_{mn}=\pm 1}
\bar{A}_{mn} \bar{B}_{mn} A_{mn} B_{mn}=
\\
\fl\nn
=1+S^2_{mn}a_{mn}b_{mn}
+S^2_{mn}(\lambda_1\,\bar{a}_{m-1n}
+\lambda_2\,\bar{b}_{mn-1})(a_{mn}+b_{mn})
+S^2_{mn}\lambda_1\lambda_2\,\bar{a}_{m-1n} \bar{b}_{mn-1}\\
 \nn
+\,S_{mn}^{4}\lambda_1\lambda_2a_{mn}b_{mn}
\bar{a}_{m-1n} \bar{b}_{mn-1}\\
 \fl
=\exp\Big[\,S^2_{mn}\Big(a_{mn}b_{mn}
+(\lambda_1 \bar{a}_{m-1n}+\lambda_2 \bar{b}_{mn-1})(a_{mn}+b_{mn})
+\lambda_1\lambda_2 \bar{a}_{m-1n} \bar{b}_{mn-1}
\Big)\Big]\,.\;\;
\label{avab1a}
\ee
The even fermionic polynomial resulting under the averaging can be written
as a Gaussian exponential, as is shown in the final line. This term is
totally commuting with all other elements of the algebra and can be removed
outside from the junction. The BC spins still remain in the form of
$S_{mn}^{2}=0,1$ in (\ref{avab1a}), but the Ising degrees, $\sign\{S_{mn}\}
=\pm1$, are already effectively eliminated. After completing the above
averaging procedure at the junction at $m=L$, for given $n$, we repeat the
calculation for $m=L-1,\, \ldots,\,2,1$, and then for other values of
$n=1,2,\,\ldots,\,L$. Adding the diagonal terms from the definition of the
fermionic averaging (\ref{fact4}), the partially traced partition function
finally reads:
\bb
\fl\nn
Z&=&2^{L^2}\sTR_{\{S^2=0,1\}}
\int  \prod_{m,n=1}^L \dd \bar {a}_{mn} \dd a_{mn}\dd \bar {b}_{mn} \dd
b_{mn}
\exp \left[\Delta S^2_{mn}
+(1+\lambda'_1\,S^2_{mn}S^2_{m+1n})a_{mn}\bar{a}_{mn}
\right.
\\
\fl\nn
& &
+(1+\lambda'_2\, S^2_{mn}S^2_{mn+1}) b_{mn}\bar{b}_{mn}
+S^2_{mn}\,(\lambda_1 \bar{a}_{m-1n}
+\lambda_2 \bar{b}_{mn-1} )(a_{mn}+b_{mn})
\\[5pt]
\fl
& &\left.
+\,S^2_{mn}\,a_{mn} b_{mn}
+\,S^2_{mn}\,\lambda_1 \lambda_2 \bar{a}_{m-1n} \bar{b}_{mn-1}
\right]\,.\;\;
\label{abss1a}
\ee
The resulting integral for $Z$ in (\ref{abss1a}) is the Gaussian integral,
which includes yet the variables $S_{mn}^{2}=0,1$ as parameters. At this
stage, it easy to recognize that the 2D Ising model is solvable, since in
this case $S_{mn}^{2}=1$ at all sites. The partition function $Z$ is then
given by a Gaussian fermionic integral, which can be readily evaluated by
passing to the momentum space for fermions \cite{ple85dok,ple85tmp}. This
results the Onsager's expressions for $Z$ and $-\beta F=\ln Z$.  In the BC
model case, it remains yet to  eliminate $S_{mn}^{2}=0,1$ degrees of
freedom in the above expression (\ref{abss1a}) for $Z$.\\
The trace over $S_{mn}^{2}=0,1$ can be performed in (\ref{abss1a}) after we
manage to decouple the  variables in terms including
$S_{mn}^{2}S_{m+1n}^{2}$ and $S_{mn}^{2}S_{mn+1}^{2}$. Several methods are
possible. A one way is to introduce another auxiliary set of Grassmann link
variables, similarly to what we previously did to decouple the factors
$S_{mn}S_{m+1n}$ and $S_{mn} S_{mn+1}$ in (\ref{fact1}).
It is possible however to avoid the introduction of the new fields by
using instead the following rescaling of the fermionic variables
under the integral: $a_{mn} \to a_{mn}/S_{mn}^{2},\,b_{mn}\to
b_{mn}/S_{mn}^{2}$.  Respectively, to preserve the integral invariant, one
has to rescale the differentials:  $da_{mn}\to S_{mn}^{2}da_{mn},\,
db_{mn}\to S_{mn}^{2}db_{mn}$. This may be viewed, in principle, as a kind
of a change of variables in a fermionic integral, which leaves the integral
invariant, as it follows from the basic rules of integration. The variable
$S_{mn}^{2}$ then disappears in some places inside the exponential and
appears in the others, the terms with $S_{mn}^{2}S_{m+1n}^{2}$ and
$S_{mn}^{2} S_{mn+1}^{2}$ being decoupled.  Also, the resulting
seemingly singular expressions like $S_{mn}^{2} \exp(a_{mn} \bar a_{mn}/
S_{mn}^{2})$ are to be understood as $S_{mn}^{2} \exp(a_{mn} \bar a_{mn}/
S_{mn}^{2}) =S_{mn}^{2}(1+a_{mn}\bar a_{mn}/ S_{mn}^{2})=S_{mn}^{2}
+a_{mn}\bar a_{mn}$. Finally, after shifting some indices in the sums,
we obtain:
\bb
\nn\fl
Z &=& 2^{L^2}\sTR_{\{S^2=0,1\}}
\int  \prod_{m,n=1}^L \dd \bar {a}_{mn} \dd a_{mn}\dd \bar {b}_{mn} \dd
b_{mn} (S^2_{mn}+a_{mn}\bar{a}_{mn})(S^2_{mn}+b_{mn}\bar{b}_{mn})
\\
\nn\fl
& \times& \exp \left[ \Delta S^2_{mn}
+ S^2_{mn}\left(\lambda'_1\, a_{m-1n}\,\bar{a}_{m-1n}
+\lambda'_2\, b_{mn-1} \bar{b}_{mn-1}
+\lambda_1 \lambda_2\,\bar{a}_{m-1n}\bar{b}_{mn-1}\right)
\right]
\\*[5pt]\fl
&\times& \exp \left[a_{mn}b_{mn}+(\lambda_1 \bar{a}_{m-1n}+\lambda_2
\bar{b}_{mn-1})(a_{mn}+b_{mn})\right]\,.\;\;
\label{bcint1}
\ee
In this expression, we can already locally perform the sum over
$S_{mn}^{2}=0,1$ at each site. The rules like (\ref{fuy3})-(\ref{fuy6})
are to be taken into account in order not to count twice the contribution
of $S_{mn}^{2}=0$ states. By averaging the part of the product explicitly
depending on $S_{mn}^{2}=0,1$, we obtain:
\bb
\nn\fl
\sum_{ \{ S^2_{mn}=0,1 \} }
\Big\{\,2^{\,S^2_{mn}}
\Big[(S^2_{mn}+a_{mn}\bar{a}_{mn})(S^2_{mn}+b_{mn}\bar{b}_{mn})
\Big]\,
\\ \nn
\times \exp \Big[S^2_{mn}\Big(\Delta+\lambda'_1a_{m-1n} \bar{a}_{m-1n}
+\lambda'_2 b_{mn-1} \bar{b}_{mn-1}
+\lambda_1 \lambda_2 \bar{a}_{m-1n}\bar{b}_{mn-1}\Big)
\Big]\Big\}
\\
=a_{mn} \bar{a}_{mn} b_{mn} \bar{b}_{mn}
+ 2 e^\Delta e^{G_{mn}}, \;\;
\label{Trrho}
\ee
where $G_{mn}$ in the exponential in final line stands for the local part
of the action resulting at the Ising site with $S_{mn}^{2} =1$. The first
term in final line is the one produced at dilute site with $S_{mn}^{2}=0$.
The explicit expression for $G_{mn}$ reads:
\bb\fl
G_{mn}=
a_{mn}\bar{a}_{mn}+b_{mn}\bar{b}_{mn}+\lambda_1 \lambda_2\,
\bar{a}_{m-1n} \bar{b}_{mn-1}
+\lambda'_1\, a_{m-1n} \bar{a}_{m-1n}+\lambda'_2\,
b_{mn-1} \bar{b}_{mn-1}.\;\;
\label{bcint3}
\ee
The result of the averaging from (\ref{Trrho}) can as well be written as
a unique exponential taking into account the nilpotent property of
fermions:
\bb
\fl\nn
a_{mn} \bar{a}_{mn} b_{mn} \bar{b}_{mn}
+2\,e^\Delta e^{G_{mn}}
=2 e^\Delta e^{G_{mn}}
\left(1+\frac{1}{2}\,a_{mn} \bar{a}_{mn} b_{mn}
\bar{b}_{mn}\, e^{-\Delta-G_{mn}} \right)
\\ \nn
=2 e^\Delta \exp \left(G_{mn}
+\frac{1}{2} a_{mn} \bar{a}_{mn} b_{mn}\bar{b}_{mn}
e^{-\Delta-G_{mn}}  \right)
\\
=2 e^\Delta \exp \left(G_{mn}
+\frac{1}{2}e^{-\Delta} a_{mn} \bar{a}_{mn} b_{mn}\bar{b}_{mn}
e^{-G'_{mn}}\right)\,.\;\;
\label{bcint4}
\ee
In the final expression, we assume the local action $G_{mn}$ to be replaced
by its reduced version $G_{mn}'$, since the prefactor $a_{mn}\bar{a}_{mn}
b_{mn} \bar{b}_{mn}$ annihilates the two first terms of $G_{mn}$ in the
exponential. The reduced action reads:
\bb
\fl
G'_{mn} =\lambda'_1\, a_{m-1n} \bar{a}_{m-1n}
+\lambda'_2\,b_{mn-1} \bar{b}_{mn-1} +\lambda_1 \lambda_2\,
\bar{a}_{m-1n} \bar{b}_{mn-1}\,.\;\;
\label{bcint4a}
\ee
Substituting this result into (\ref{bcint1}) and shifting the $mn$ index
in some of the diagonal terms of the resulting combined action,
we obtain:
\bb
\nn\fl
Z=2^{L^2} e^{L^{2}\Delta}
\int \prod\limits_{m=1}^{L}\prod\limits_{n=1}^{L}
d\bar{a}_{mn}da_{mn}d\bar{b}_{mn}db_{mn}
\exp\Big\{\sum\limits_{m=1}^{L}
\sum\limits_{n=1}^{L}
\Big[\,(1+\lambda_1')\,a_{mn}\bar{a}_{mn}
\\ \nn \fl
+(1+\lambda_2')\,b_{mn}\bar{b}_{mn}
+a_{mn}b_{mn} +(\lambda_1\bar{a}_{m-1n} +\lambda_2\bar{b}_{mn-1})
(a_{mn} +b_{mn})
\\ \nn \fl
+\lambda_1\lambda_2\,\bar{a}_{m-1n}\bar{b}_{mn-1}
+\bar{g}_0 \;a_{mn}\bar{a}_{mn}b_{mn}\bar{b}_{mn}\,
\exp\,(-\lambda_1'\,a_{m-1n}\bar{a}_{m-1n}
-\lambda_2'\, b_{mn-1}\bar{b}_{mn-1}
\\ \fl
-\lambda_1\lambda_2\,
\bar{a}_{m-1n}\bar{b}_{mn-1})\Big]\Big\}\,, \;\;\;\;\;
\bar g_0=e^{-\Delta}/2\,,\;\;
\label{bcint5}
\ee
with $-\Delta =+\beta\Delta_0$. This is already a purely fermionic integral
for $Z$, the spin degrees of freedom being completely eliminated.  The
interaction part of the action is introduced with coupling constant
$\bar{g}_0\propto e^{\beta\Delta_0}$, which depends on $\Delta_0$. To
simplify the comparison with the 2D Ising model, and for other needs, we
now rescale some of the Grassmann variables under the integral using
the following transformation:
\bb
(1+\lambda'_1)\,\bar{a}_{mn} \rightarrow \bar{a}_{mn},\;\;\;\;
(1+\lambda'_2)\,\bar{b}_{mn} \rightarrow \bar{b}_{mn}.
\;\;\;\label{bcint5b}
\ee
The corresponding differentials are to be rescaled with inverse factors.
In this way, we obtain the final result of this subsection:
\bb
\fl\nn
Z=(2e^{\Delta}\cosh K_1 \cosh K_2)^{L^2}
\int \prod\limits_{m=1}^{L}\prod\limits_{n=1}^{L}
d\bar{a}_{mn}da_{mn}d\bar{b}_{mn}db_{mn}\exp\Big\{\sum\limits_{m=1}^{L}
\sum\limits_{n=1}^{L}
\\ \nn \fl
\Big[\,a_{mn}\bar{a}_{mn} +b_{mn}\bar{b}_{mn}
+a_{mn}b_{mn} +(t_1\bar{a}_{m-1n} +t_2\bar{b}_{mn-1})
(a_{mn} +b_{mn})
+t_1t_2\,\bar{a}_{m-1n}\bar{b}_{mn-1}
\\  \fl
+\;g_0\;a_{mn}\bar{a}_{mn}b_{mn}\bar{b}_{mn}\,
\exp\,(-\gamma_1a_{m-1n}\bar{a}_{m-1n}
-\gamma_2b_{mn-1}\bar{b}_{mn-1}
-t_1t_2\,\bar{a}_{m-1n}\bar{b}_{mn-1})\Big]\Big\}\,,
\label{PFfinal}
\ee
where $t_i =\tanh K_i$ and we have introduced the following constants:
\bb
\fl
g_0=\frac{e^{-\Delta}}{2 \cosh K_1 \cosh K_2},\;\;\;
\gamma_{i}=1-\frac{1}{\cosh K_{i}}=1-\sqrt{1-t^2_{i}}.\;\;\;
\;\;\;\label{bcint8}
\ee
In a compact form, the equation (\ref{PFfinal}) reads:
\bb
\fl
Z =(2e^{\Delta}\cosh K_1 \cosh K_2)^{L^2}\int D\bar a
DaD\bar b Db \;\;\exp(\Sising +\Sint)\,.\;\;\;
\label{bcint8a}
\ee
Note that the fermionic integrals (\ref{bcint5}) and (\ref{PFfinal}) for
the Blume-Capel partition function $Z$ are still the exact expressions.
They both are equivalent to each other and to (\ref{PFspin}). The above
correspondence is exact even for finite lattices, provided we assume free
boundary conditions both for spins and fermions. \footnote{ \ The exact
fermionic representation for $Z$ for a finite lattice with periodic
boundary conditions for spin variables in both directions also can be
derived. The result will be the sum of four fermionic integrals like
(\ref{bcint5}) and (\ref{PFfinal}), with periodic-aperiodic boundary
conditions for fermions, in analogy to the case of the 2D Ising model on
a torus (cf. \cite{ple85tmp,liaw99,wuhu02}).} We can recognize in
(\ref{PFfinal}) and (\ref{bcint8a}) the Ising action, which is simply
the Gaussian part of the total action (cf. \cite{ple85dok,ple85tmp}):
\bb
\fl\nn \Sising = \sum_{m,n=1}^L a_{mn}\bar{a}_{mn}
+b_{mn}\bar{b}_{mn}+a_{mn}b_{mn}\\ \label{Sising} +(t_1\bar{a}_{m-1n}
+t_2\bar{b}_{mn-1})(a_{mn}+b_{mn}) +t_1t_2 \bar{a}_{m-1n}
\bar{b}_{mn-1},\;\; \label{SIsing1}
\ee
and the non-Gaussian interaction part of the total action, which is
a polynomial of degree 8 in Grassmann variables (after expanding the
exponential):
\bb
\fl
\label{Sint} \Sint =g_0 \sum_{m,n=1}^L
a_{mn}\bar{a}_{mn}b_{mn}\bar{b}_{mn} e^{
-\gamma_1a_{m-1n}\bar{a}_{m-1n}-\gamma_2b_{mn-1}
\bar{b}_{mn-1}-t_1t_2\,\bar{a}_{m-1n}\bar{b}_{mn-1}}.\;\;
\label{Sint1}
\ee
The BC model differs from the Ising model by the interaction term
(\ref{Sint1}) in the total action, which is not quadratic. Therefore the
BC model is not solvable in the sense of free fermions, as distinct from
the pure 2D Ising model.\\
It may be still of interest to try to recognize the structure of the phase
diagram of the BC model directly from the fermionic integrals
(\ref{bcint5})-(\ref{Sint1}) before actual calculation. The interaction is
introduced in the above BC integral (\ref{PFfinal}) for $Z$ with the
coupling constant $g_0\propto \exp[\beta(\Delta_0-J_1 -J_2)]$.  In the
limit $\Delta \rightarrow \infty$ (or $\Delta_0\to -\infty$), which
corresponds to $g_0=0$, the gap between the two degenerate states $S=\pm 1$
and the singlet state $S=0$ becomes infinitely large and the model reduces
effectively to the 2D Ising model. For $\Delta_0$ finite, the coupling
constant $g_0$ is finite and the presence of the vacancy states becomes
possible. The coupling constant $g_0$ increases as
the number of the vacancies in a typical configuration of a system
increases, with increasing $\Delta_0$. At zero temperature, on the other
hand, as $\beta\to+\infty$, we find $g_0=0$ for $\Delta_0<J_1+J_2$, which
corresponds, again, to Ising ground state, while for $\Delta_0>J_1+J_2$ we
have $g_0\to+\infty$, which will mean that the ground sate is empty (all
sites are occupied by vacancies). These features at $T=0$ can be
readily guessed already from the form of the original Hamiltonian
(\ref{ham1a}).  A more sophisticated analysis of the integral
(\ref{PFfinal}) will be needed to define the precise form of the critical
line and to locate the tricritical point at that line with increasing
dilution.\\ In the following, for simplification, we will only consider the
isotropic coupling case, with $K_1=K_2=K$, $t_1=t_2=t$ and
$\gamma_1=\gamma_2 =\gamma$.

\subsection{Partial bosonization}

The previous action contains two pairs of Grassmann variables per site.
This can not be reduced to a one pair (minimal action) unlike the Ising
model, where half of the variables are irrelevant in the sense they
do not contribute to the critical behaviour and can be integrated out
already at lattice level. The point is that the reduced Ising action with
two variables per site readily admits QFT interpretation and simplifies the
analysis in the momentum space \cite{ple98,ple98fl, dritz89, ple95amm}. In
the BC case, the two pairs of fermions are coupled together by Eq.
(\ref{Sint}), preventing a direct integration over extra variables like
$a_{mn},b_{mn}$.  However, as we will see in the following, it is still
possible to recover the minimal Ising like action with a one pair of
fermions per site using auxiliary bosonic variables. In the interaction
part of the action (\ref{Sint}), it is indeed tempting to replace the
products $a_{mn}\bar a_{mn}$ and $b_{mn}\bar b_{mn}$, which are formally
looking similar to occupation number operators, or local densities, by the
new commuting variables as follows:

\bb
\eta_{mn}=a_{mn}\bar a_{mn}\,,\;\;\;\; \tau_{mn} =b_{mn}\bar
b_{mn}\,,\;\;\;\;\; \eta_{mn}^2=\tau_{mn}^{2}=0\,.\;\;
\label{eta1a}
\ee
These new variables $\eta_{mn},\tau_{mn}$ are nilpotent (as Grassmann
variables) but purely commuting: that is why we will abusively call them
(hard core) ``bosons". In the following, we will add also one more pair of
commuting nilpotent fields $\bar{\eta}_{mn}, \bar{\tau}_{mn}$, to put
integrals into a more symmetric form. The identities like (\ref{eta1a})
are rather to be understood in the sense of correspondence, to be realized
by a properly introduced delta-functions (Dirac distributions). This
eventually allows us to reduce the degree of polynomials in Grassmann
variables by a factor 2 each time the replacement like (\ref{eta1a}) is
performed, even if terms like $\bar{a}_{m-1n}\bar{b}_{mn-1}$ in
(\ref{Sint}) can not be replaced. We will
see below that, in principle, we can write down an action containing a one
pair of Grassmann variables and a one pair of bosonic ones per site. To do
so, we introduce the following Dirac distribution for any polynomial
function $f$ of nilpotent variables like $a_{mn}\bar a_{mn}$ or $b_{mn}
\bar b_{mn}$:
\bb
\nn\fl
f(a_{mn}\bar a_{mn})=\int \dd \eta_{mn}\dd \bar\eta_{mn}
f(\eta_{mn})\exp \left [
\bar\eta_{mn}(\eta_{mn}+a_{mn}\bar a_{mn})\right ],\;\;
\\ \fl
f(b_{mn}\bar b_{mn})=\int \dd \tau_{mn} \dd \bar\tau_{mn}
f(\tau_{mn})\exp \left [
\bar\tau_{mn}(\tau_{mn}+b_{mn}\bar b_{mn})\right ].\;\;
\label{dirac1}
\ee
We assume a natural definition of the integral for commuting nilpotent
variables with the following rules of integration (similar rules are
assumed for $\bar{\eta}_{mn}, \bar{\tau}_{mn}$):
\bb
\int d\eta_{mn}\,(1, \eta_{mn}) =(0,1)\,,\;\;\;
\int d\tau_{mn}\,(1, \tau_{mn}) =(0,1)\,. \;\;\;
\label{etaint1}
\ee
For application of the rules like (\ref{etaint1}) in the QFT context also
see \cite{palumbo97}.  Applying (\ref{dirac1}) directly in (\ref{PFfinal}),
we obtain the integral with the following expression for the
action:

\bb
\nn\fl
\action=\sum_{m,n}
\Big [
a_{mn}\bar a_{mn}+b_{mn}\bar b_{mn}+t^2\bar a_{m-1n}\bar b_{mn-1}
+ a_{mn}b_{mn}+t(\bar a_{m-1n}+\bar b_{mn-1})(a_{mn}+b_{mn})
\\ \nn
+\,g_0\,\eta_{mn}\tau_{mn}
[1-\gamma(\eta_{m-1n}+\tau_{mn-1})
+\gamma^2\eta_{m-1n}\tau_{mn-1}
-t^2\bar a_{m-1n}\bar b_{mn-1}]
\\ \,\label{act1aa}
+\bar\eta_{mn}(\eta_{mn}+a_{mn}\bar a_{mn})+
\bar\tau_{mn}(\tau_{mn}+b_{mn}\bar b_{mn})\Big]\,.\;\;
\ee
We can now integrate over the $a_{mn}$'s and $b_{mn}$'s, and replace
formally, for convenience, the variables $\bar a_{mn}$ by $c_{mn}$ and
$\bar b_{mn}$ by $-\bar c_{mn}$ in the remaining integral.
We obtain:
\bb
\nn\fl
\action=\sum_{mn=1}^L
\Big\{ c_{mn}\bar c_{mn}(1+\bar\tau_{mn})(1+\bar\eta_{mn})
+\bar\eta_{mn}\eta_{mn}+\bar\tau_{mn}\tau_{mn}
\\ \nn\fl
+[c_{mn}(1+\bar\eta_{mn})-\bar c_{mn}(1+\bar \tau_{mn})]
t(c_{m-1n}+\bar c_{mn-1})
-t^2c_{m-1n}\bar c_{mn-1}
\\ \fl
+g_0\,\eta_{mn}\tau_{mn}
\Big [1-\gamma(\eta_{m-1n}+\tau_{mn-1})
+\gamma^2\eta_{m-1n}\tau_{mn-1}
-t^2c_{m-1n}\bar c_{mn-1}\Big]\Big\}\,.
\label{act1bb}
\ee
The advantage is that now there are only two fermionic variables per
site, which is suitable for the QFT interpretation \cite{ple98,ple95amm}.
Note that the integral associated with the action (\ref{act1bb}) will still
be the exact expression for $Z$. The number of the fermionic variables
being reduced, the next operation is to try to integrate out, whenever
possible, the auxiliary bosonic fields from action (\ref{act1bb}). In fact,
we can further integrate over one pair of bosonic variables, for example
$\tau_{mn},\,\bar\tau_{mn}$, using the integration rules like
(\ref{dirac1}), since
\bb
\nn\fl
\int d\tau_{mn} d\bar\tau_{mn}\,f(\tau_{mn})\,
\exp[\bar\tau_{mn}\big(\tau_{mn}-t(c_{m-1n}-\bar c_{mn-1})\bar c_{mn}
+c_{mn}\bar c_{mn}(1+\bar \eta_{mn})\big)]
\\
=\,f\,[-t(c_{m-1n}-\bar c_{mn-1})\bar c_{mn}
+c_{mn}\bar c_{mn}(1+\bar \eta_{mn})]\,.\;\;
\label{act1ac}
\ee
There $f(\tau_{mn})$ may be any function of nilpotent variable $\tau_{mn}$.
We could also have chosen to integrate over the $\eta_{mn}, \bar \eta_{mn}$
instead. Integrating over $\tau_{mn},\bar \tau_{mn}$ according
to (\ref{act1ac}), we finally obtain the reduced integral with the local
action:
\bb
\nn\fl
\action =c_{mn}\bar c_{mn}+t(c_{mn}+\bar c_{mn})(c_{m-1n}-\bar c_{mn-1})
-t^2c_{m-1n}\bar c_{mn-1}
\\ \nn
+\bar\eta_{mn}\eta_{mn}
+\bar\eta_{mn}\Big [\bar c_{mn}-t(c_{m-1n}-\bar c_{mn-1})
\Big] c_{mn}
\\ \nn
+g_0\,\eta_{mn}Q_{mn}\Big [1-\gamma(\eta_{m-1n}+Q_{mn-1})
\\
+\gamma^2\eta_{m-1n}Q_{mn-1}
+t^2c_{m-1n}\bar c_{mn-1}\Big],\;\;
\label{act1fin}
\ee
with
\bb
Q_{mn}=[c_{mn}(1+\bar\eta_{mn})-t(c_{m-1n}-\bar c_{mn-1})]
\bar c_{mn}\,.\;\;
\label{act1aq}
\ee
It is easy to recognize in the first line of (\ref{act1fin}) the minimal
local action for the pure Ising model \cite{ple98,ple95amm} with one pair
of Grassmann variables per site:
\bb
\fl
\Sising=c_{mn}\bar c_{mn}
+t(c_{mn}+\bar c_{mn})(c_{m-1n}-\bar c_{mn-1})
-t^2c_{m-1n}\bar c_{mn-1}.\;\;
\label{SIsing2}
\ee
This is the same action that follows by integrating $a_{mn},b_{mn}$ from
(\ref{SIsing1}). The rest of the action describes the interaction between
fermions and bosons:
\bb
\fl \nn
\label{SIint1}
\Sint=\bar\eta_{mn}\eta_{mn}+\bar\eta_{mn} c_{mn}\Big [
\bar c_{mn} +t(c_{m-1n}-\bar c_{mn-1})
\Big ]
\\ \nn
+\,g_0\,\eta_{mn}Q_{mn}\Big [1-\gamma(\eta_{m-1n}+Q_{mn-1})
\\
+\,\gamma^2\eta_{m-1n}Q_{mn-1}+t^2c_{m-1n}\bar c_{mn-1}\Big ].
\ee
It is easy to check that at $g_0=0$ the boson variables can be integrated
out in the action (\ref{act1fin}). This may be less simple task for finite
values of coupling constant $g_0\neq 0$. In the next section, we will apply
approximations in order to eliminate completely the auxiliary commuting
nilpotent fields from the action, and will make use of a more symmetric
form of the integration over the bosonic fields, first over
$\bar{\eta}_{mn},\bar{\tau}_{mn}$, then
over $\eta_{mn},\tau_{mn}$.

We would like to end this section commenting the previous exact results.
We finally obtained a lattice field theory with action (\ref{act1fin})
containing the same number of ``fermions" ($c$, $\bar c$) and ``bosons"
($\eta$,~$\bar\eta$).  Physically, this means that it is indeed possible to
describe the system with fermionic variables for the states $S=\pm 1$ and
bosonic ones for the third state $S=0$. In the limit $\Delta_0 \rightarrow
-\infty$, the system is completely described in terms of fermions. While
with $\Delta_0$ increasing to finite values, an interaction between
fermions and bosons is added. Beyond a value $\Delta_{0 t}$,
fermions form bosonic pairs: in the limit $\Delta_0 \rightarrow + \infty$,
all fermions condense into bosons, leading to a purely bosonic system.
In this interpretation, the tricritical point may be expected to be seen
as a particular point on the critical line where the interaction is such
that an additional symmetry between fermions and bosons appears. This might
correspond to supersymmetry appearing in the conformal field theory
describing tricritical Ising model. To our knowledge there is no evidence
of supersymmetry derived directly from a lattice model: the exact lattice
action (\ref{act1fin}) could be a good way to see how super-symmetry may
emerge from a lattice model. Of course all we said so far is only
speculative: we are currently studying it in more detail, to confirm or
infirm this hypothesis.

\section{Effective action in the continuum limit}

In the Ising model case, the fermionic action on the lattice is quadratic
and the corresponding Grassmann integral can actually be computed exactly
by transformation into the momentum space for fermions by means of the
Fourier substitution.  The situation for the 2D BC model is less simple, as
there is a non-Gaussian interaction part in action (\ref{act1bb}), alias
(\ref{act1fin}), which contains terms of order up to 8th in fermions. The
Grassmann integral leading to the partition function can no longer be
computed directly, as in the Ising model case, by a simple Fourier
substitution. In this sense the 2D BC model is not integrable. However it
is still possible to extract physical information by taking the continuous
limit of the BC lattice action like (\ref{PFfinal}), or (\ref{act1bb}),
and analyzing it using tools from quantum field theory.

\subsection{Effective 2nd order fermionic field theory}

We would like to obtain an effective purely fermionic theory for the BC
model up to order 2 in momentum $\vk\,$ from the previous calculations,
with two variables per site, to analyze the critical behavior of
the model. In the Ising case, the critical behavior is given, in the
continuous limit, by a massless Majorana theory that follows from
two-variable action. In the following, we will see how to compute the mass
of the BC model in its effective Gaussian part. The condition of the zero
effective mass will give already the critical line in the $(T,\Delta_0)$
plane for BC model. For the location of a tricritical point on that line
one needs more complicated analysis, taking into account the stability of
the kinetic part of the action, which is in turn affected by the presence
of the interaction. In the infrared limit, the spectrum is given by
expanding the  effective action, or rather the correspondent partial
integral $Z_{\vk}$ in $Z=\prod_{\vk}Z_{\vk}$, up to the second order in the
momentum $\vk$. The coefficient $\lambda$ in front of the term
$\lambda\vk^2$ in the basic factor $Z_{\vk}$ of $Z$ is what we call the
{\it stiffness} parameter of the model. It dominates all contributions from
the kinetic part of the action. In the Ising model case, the stiffness
coefficient is always a strictly positive coefficient. In this case, the
only singularity in the spectrum follows from the condition of vanishing
mass, resulting the Ising critical point. Here in the BC model, we will
show that the effective stiffness coefficient can also vanish at a special
point at the critical line in $(T,\Delta_0)$ plane, rendering the spectrum
unstable and changing the nature of the singularity.  This happens for
large enough $g_0$, as $\Delta_0$ increases. We intend to identify the
above singular point as an evidence for the appearance of a tricritical
point, together with a segment of the first-order transition line, at the
BC phase diagram at sufficiently strong dilution. In order to be able to
perform the QFT analysis of the above kind, we ought to eliminate the
bosonic nilpotent fields from the action, being interested merely in the
low-momentum (small $\vk$) sector of the theory, and making reasonable
approximations whenever necessary.\\
This program also implies a more symmetric way of integration over the
nilpotent fields. Instead of integrating over the variables $\tau_{mn}$
and $\bar\tau_{mn}$ as in Eq. (\ref{act1fin}), we now proceed by
integrating first over $\bar\eta_{mn}$ and $\bar\tau_{mn}$ in
Eq.~(\ref{act1bb}), making use of the definition of the integral. This
results the reduced integral with a new action:
\bb
\nn\fl
Z_{}=(2e^{\Delta}\cosh^2K)^{L^2}\int \prod_{m,n} \dd \bar c_{mn} \dd c_{mn}
\dd \eta_{mn} \dd \tau_{mn} \left [ c_{mn}\bar
c_{mn}+\eta_{mn}q_{mn}+\tau_{mn}\bar q_{mn} +\eta_{mn}\tau_{mn} \right  ]\\
\times \exp(\Sising+\Sint)\,,\;\;
\label{aci1ii}
\ee
where $\Sising$ is given in (\ref{SIsing2}), while
\bb
\fl
\Sint=g_0\sum_{m,n}\eta_{mn}\tau_{mn}\left
[(1-\gamma\eta_{m-1n})(1-\gamma\tau_{mn-1})
+t^2 c_{m-1n}\bar c_{mn-1}\right ],
\label{aci2ii}
\ee
and
\bb\fl\nn
\bar q_{mn}=c_{mn}\bar c_{mn} +tc_{mn}(c_{m-1n}-\bar c_{mn-1})
=c_{mn}[\bar c_{mn} +t(c_{m-1n}-\bar c_{mn-1})]\,,
\\
\fl
q_{mn}=c_{mn}\bar c_{mn} +t\bar c_{mn}(c_{m-1n}-\bar c_{mn-1})
=[c_{mn} -t(c_{m-1n}-\bar c_{mn-1})]\bar{c}_{mn}\,.\;\;
\label{aqq1}\;\;
\ee
It is also useful to note that $q_{mn}^{2} =\bar{q}_{mn}^{2}=0$, and
$q_{mn}\bar q_{mn}=0$. The free-fermion Ising part of the action
$\Sising$ in (\ref{aci1ii}) at this stage remains unchanged and is given by
the standard expression (\ref{SIsing2}). The above integral (\ref{aci1ii})
includes as well the product of quadratic polynomial terms like $c_{mn}
\bar c_{mn}+\eta_{mn}q_{mn}+\tau_{mn}\bar q_{mn} +\eta_{mn} \tau_{mn}$,
which can not be written as a single exponential. However, when
integrating over the remaining variables $\eta_{mn}$ and $\tau_{mn}$, it is
easy to realize that these polynomial terms \textit{roughly} impose the
following substitution rules in the action $\Sint$:
\bb
\eta_{mn}\tau_{mn}\rightarrow c_{mn} \bar c_{mn}\,,\;\;\;
\eta_{mn}\rightarrow \bar q_{mn}\,,\;\;\;
\tau_{mn}\rightarrow q_{mn}\,.\;\;\;
\label{eta1aa}
\ee
In a sense, the above rules can be considered as an operation of
approximate Dirac delta functions on the variables $\eta_{mn}$ and
$\tau_{mn}$, replacing them by fermions. These rules of correspondence
though are {\em not} unreservedly exact: when expanding the exponential
of $\Sint$ into a series, the terms will appear that couple to each other
to give $c_{mn}\bar c_{mn}$ but not $q_{mn} \bar q_{mn}=0$ as is given by
the above substitution rules.  For example, terms such as
\bb
(g_0\eta_{m+1n}\tau_{m+1n}\gamma\eta_{mn})
\times(g_0\eta_{mn+1}\tau_{mn+1}\gamma\tau_{mn}),
\label{eta1bb}
\ee
instead of vanishing, lead to a contribution in the effective action
equal to
\bb
g_0^2\gamma^2c_{mn}\bar c_{mn}c_{m+1n}\bar c_{m+1n}
c_{mn+1}\bar c_{mn+1}.
\label{example}
\ee
Therefore there are more terms in the final effective action
$\action_{\mathrm{eff}} (c,\bar{c})$ than in the one resulting from the
above substitution rules. However, we would like to apply  approximations
to the term with interaction, and the higher order corrections of the above
kind can be neglected within this scheme anyhow.
From an effective action that follows from (\ref{aci1ii}), we intend to
obtain the basic momentum-space factor $Z_{\vk}$ of $Z$ up to order 2 in
momentum $\vk$, in order to study the stability of the free fermion
spectrum. In the pure Ising model at criticality, the factor
$Z_{\vk}$ gives basically a $(\underline{m}^2 +\vk^2)$ contribution to the
partition function and free energy at small momenta, with mass
$\underline{m}^2=0$ at the critical point \cite{ple98,dritz89}. In fact,
there is also the {\em stiffness} coefficient $\lambda$ in front of $\vk^2$
in this  term, $\vk^2\to \lambda \vk^2$. In the Ising case, this stiffness
coefficient is non-singular at the critical point and can be fixed simply
by its finite value at the critical temperature. In the BC case, however,
we have a line of critical points as $\Delta_0$ varies from negative to
positive values.  Respectively, the stiffness coefficient $\lambda
=\lambda( \Delta_0)$ also varies with a variation of the chemical potential
$\Delta_0$ along the critical line. The point is that in the BC case the
effective stiffness coefficient vanishes at some position at the critical
line, for a sufficiently strong dilution, which may eventually be
identified as the tricritical point of the  BC model. In what follows, we
apply the Hartree-Fock-Bogoliubov (HFB) approximating scheme
\cite{thouless72, mattuck92, bogoliubov07} in the momentum space in order
to gain a modification of the above Ising like behaviour provided by the
presence of the interaction in the BC case. In essence, the  HFB decouples
the four-fermion interaction into few Gaussian terms added to the basic
action.
\footnote{ \ Let us remember that the interaction terms in BC model appear
solely due to the presence of the dilute (vacancy) sites.  Respectively,
the strength of interaction (the coupling parameter $g_0$) increases with
increasing rate of dilution, with variation of the chemical potential
$\Delta_0$. The corrections with $g_0$ may thus appear in the mass term and
the stiffness coefficients of the BC effective action within mean-field HFB
analysis. In fact, as we shall see below, the relevant $g_0$ correction to
the mass at the Gaussian (free-fermion) level already follows when we
extract the effective action from (\ref{aci1ii}) and (\ref{eta1aa}), see
(\ref{Seff1}), while the kinetic corrections, that at lattice level may be
attributed to the correlations of the Ising degrees and vacancies at the
same and neighbouring sites, are to be extracted self-consistently within
HFB scheme from the residual interaction in the effective action.} This
also assumes a self-consistent calculation of the corrections which modify
the parameters in the mass term and the kinetic part of the action, and
eventually modify the stiffness coefficient, due to the HFB decoupling of
the interaction.\\
Among the terms that contribute in $Z_{\vk}$ to the second order in
momentum are in any case those coming from the kinetic part of the
free-fermion quadratic piece of the action, cf. Eq.~(\ref{SIsing2}).
In the
continuous limit, with $c_{m-1n}\to c-\partial_x c,\; \bar{c}_{mn-1}\to
\bar{c} -\partial_y \bar{c}$, these terms are combinations of
$c\,\partial_x c$ or $\bar c\,\partial_x c,\; \bar c\,\partial_y \bar c$ or
$c\,\partial_y\bar c$. From the above rules (\ref{eta1aa}), we expect that
the effective action will contain as well quartic contributions such as $c
\bar c\, \partial_{i}c\, \partial_j \bar c$, with $i,j=x,y$, at the lowest
order.  This term is degree 4 in Grassmann variables and 2 in derivatives.
The expansion of the exponential of such terms will give corrective
coefficients to the $\vk^2$ behaviour, and may thus change the order of the
transition if the renormalized stiffness vanishes. We also have to consider
not only the direct substitution of the variables with the rules given
above, but also the possible correction terms like (\ref{example}) that may
contribute to the stiffness. We should also drop terms which contain a
ratio of number of derivatives to the number of Grassmann variables higher
strictly than 1/2 as their effect is expected to provide next-order
corrections within the basic approximation scheme outlined above. After
some algebra, these following terms contribute to the effective action:

\bb
\nn\fl
\action_{\mathrm{effective}}
=\Sising+g_0\sum_{m,n}c_{mn}\bar c_{mn}\left[
(1-\gamma\bar q_{m-1n})(1-\gamma q_{mn-1})
+t^2 c_{m-1n}\bar c_{mn-1}\right ]
\\ \label{Seff1}
+g_0^2\gamma^2\sum_{m,n} c_{mn}\bar c_{mn}c_{m+1n}\bar c_{m+1n}
c_{mn+1}\bar c_{mn+1} +\ldots\;. \;\;
\ee
The above effective action define some lattice fermionic theory with
interaction. We keep in mind to analyze it further on in the momentum
space at low momenta, which corresponds to the continuum-limit
interpretation of the model.

\subsection{Continuum limit}

In the continuous-limit interpretation of the above action, we replace
$c_{mn}$ by $c=c(x,y)$ and $\bar c_{mn}$ by $\bar c=\bar c(x,y)$, assuming
as well the substitution rules like $c_{m-1n}=c-\partial_x c$ and $c_{mn-1}
=c-\partial_y c$. After a Fourier transformation of the fields, this
corresponds to the low-momenta sector of the exact lattice theory around
the origin ${\vk}=0$. In particular, we put $q_{mn}\rightarrow q=c \bar
c\,(1-t) +t(\partial_x c -\partial_y \bar c)\,\bar{c}$ and $\bar q_{mn}
\rightarrow \bar q=c\, \bar c (1-t)-tc\,(\partial_x c-\partial_y \bar c)$.
The free Ising part $\Sising$ from (\ref{Seff1}) gives simply
\bb
\label{SIsingCont} \fl
\Sising=\int \dd x \dd y
\left [
(1-2t -t^2)\, c\bar c -t(t+1)\,\bar{c}\partial_x c
+t(t+1)c\partial_y\bar c -tc\partial_x c+t\bar c
\partial_y\bar c \right].\;\;
\ee
In the above action, one
can readily distinguish the mass term and the kinetic
part, provided one assumes the QFT interpretation of the associated
integrals \footnote{ \ The Ising mass is easily seen from
(\ref{SIsingCont}) to be $\underline{m}_{\mathrm{Ising}}=1-2t-t^2$, which
must vanish at the critical point. Indeed, the condition of vanishing mass
$1-2t -t^2=0$ gives $t_c =\sqrt{2}-1$, alias $K_c =J/T_c
=\frac{1}{2}\ln(1+\sqrt{2})$, in agreement with the exact solution of this
model on a lattice. The ordered phase corresponds to negative mass, with
$t\to 1$ as $T\to 0$. The structure of the action (\ref{SIsingCont}) rather
implies the interpretation of the pure 2D Ising model in terms of the
Majorana fermions \cite{ple98,dritz89,ple95amm}. Respectively, one may pass
to the Dirac interpretation by doubling the number of fermions in the
action. Notice also that $\bar{c}\partial_{i}c =c\partial_{i}\bar{c}$
under the integral (\ref{SIsingCont}) since $\partial_i$ is a
skew-symmetric operator.}. Notice that the next order momentum term with
product $\partial_x \partial_y$ is neglected in the above action
(\ref{SIsingCont}) at the backbone of the first order $\partial_x,\,
\partial_y$ terms \footnote{ \ Despite these terms with $\partial_x$ and
$\partial_y$ are linear in momentum in the {\em action}, they contribute as
$\vk^2$ into the spectrum factor $Z_{\vk}$ of $Z$, while their product may
only contribute at the level of next order corrections to $Z_{\vk}$, as
$\underline{m}\to 0$.  }.  In the continuous limit, the last term of the BC
action (\ref{Seff1}) gives $c\bar c \partial_x c\partial_x \bar c\partial_y
c\partial_y \bar c$, which is 4$^{\mathrm{th}}$ order in derivatives and
6$^{\mathrm{th}}$ degree in Grassmann variables.  The ratio of these
numbers is 2/3 which is higher than 1/2, and therefore this term can be
discarded, as explained above. The term in factor of $g_0$ in (\ref{Seff1})
contains $\bar q_{m-1n}$ and $q_{mn-1}$ which need to be expanded up to the
order 2 in derivatives, with $\partial_{xx}\bar q=2(1-t)\partial_x
c\partial_x \bar c$, and $\partial_{yy}q =2(1-t)\partial_y c\partial_y\bar
c$. The effective action finally can be written in the continuous limit as
\fl
\bb\fl
\action_{\mathrm{eff}}
=\Sising
+\int \dd x \dd y \Big\{g_0c\bar c +g_0c\bar c
\big[ t(t+2\gamma)\partial_y c\partial_x\bar c
-\gamma(1-t)(\partial_x c\partial_x\bar c
+\partial_y c\partial_y\bar c) \big] \Big\}\;.\;\;
\label{SeffCont}
\ee
In the following, we shall use this effective action to obtain information
on the phase diagram of the BC model.

\section{Spectrum analysis and phase diagram}

In this section we analyze the critical properties of the effective action
(\ref{SeffCont}) and the low energy spectrum $Z_{\vk}$ of $Z$ in the
momentum-space representation. In particular we develop a physical argument
for the existence of a tricritical point on the phase diagram from the
above fermionic action. The critical line follows already from the
condition of the zero mass. At the tricritical point, we assume that
the effective stiffness coefficient in factor $Z_{\vk}$ also vanishes. The
Hartree-Fock-Bogoliubov (HFB) approximation scheme will be used to count
properly the effects of the interaction \cite{thouless72, mattuck92,
bogoliubov07}.

\subsection{Phase diagram}
The BC model effective action of (\ref{SeffCont}) includes the
free-fermion Gaussian part and the quartic interaction. The quadratic
(Gaussian) part of the whole action is merely formed from $\Sising$, but
the remaining interaction term in the effective action (\ref{SeffCont})
also includes quadratic term $g_0\,c\bar{c}$, which is to be added to the
Ising part of the action and will modify the Ising mass. The pure Ising
model action is shown in (\ref{SIsingCont}) above. In the continuum
limit, see (\ref{SIsingCont}), this action includes the mass term
$\underline{m}_{\,Ising}\,c\bar{c}$, with $\underline{m}_{\,Ising} =1-2t
-t^2$, and the kinetic part. The condition for the critical point in the
pure Ising case is then given by $\underline{m}_{\,Ising}=0$
\cite{ple98,dritz89,ple95amm}. In the BC case, the presence of the
Gaussian correction $g_0\,c\bar{c}$ will modify the mass term in the
effective BC action: $\underline{m}_{\,Ising} \to \underline{m}_{\,BC}
=1+g_0 -2t -t^2$, which we assume to be vanishing at the critical
line.\footnote{ \ The additive corrections that may contribute to the mass
term from the non-Gaussian part of the action (\ref{SeffCont}) are $\vk^2$
dependent and vanish as $\vk^2\to 0$. They may be neglected. In the
effective action (\ref{SeffCont}), the principal modification of the mass
term due to vacancies is realized already at the Gaussian level by the
$g_0\,c\bar{c}$ term, as it is commented above. The effect of the
non-Gaussian part in (\ref{SeffCont}) is merely that it produces
corrections to the kinetic terms, after the HFB decoupling of the
interaction.}\\
The approximations we intend to apply to tackle the
remaining quartic part of the BC action (\ref{SeffCont}) are of that kind
that we replace, in different possible ways, the two of four fermions by
variational parameters, or the effective binary averages, which are then
specified self-consistently from the resulting Gaussian action. This may be
viewed as a kind of the HFB like approximation method, which proved to be
effective in systems of quantum interacting fermions, like BCS theory of
ordinary superconductivity. This also implies that calculations are to be
performed rather in the momentum space, but not on the real lattice, or its
continuum real space version, and the correspondent symmetries are to be
taken into account properly. The application of the HFB scheme also implies
that the interaction may be not necessary weak.\\
From the explicit form of the
quartic part of the interaction in (\ref{SeffCont}), it can be seen that
the decoupling of the quartic part of $S_{int}$ produces terms which only
modify the kinetic terms in the effective action, at least, in first
approximation, with calculations being up to order $\vk^2$ in the $Z_{\vk}$
factor. This modification might be significant at strong dilution,
rendering the appearance of the tricritical point and changing the nature
of the phase transition from second to the first kind. These effects are to
be discussed in the second part of this section. In next subsection, we
consider in more detail the BC critical line in the $(T,\Delta_0)$ plane,
with dimensionless temperature $T$ and chemical potential  $\Delta_0$
normalized by the exchange energy $J$.

\subsection{Critical line}

The equation for the BC critical line we consider in this section is the
one that follows from the condition of vanishing mass,
$\underline{m}_{\mathrm{BC}} =1+g_0 -2t -t^2=0$. In a detailed form, this
equation reads:
\bb
\label{criticalline}
\tanh^2\left(\frac{1}{T}\right)+2\tanh\left(\frac{1}{T}\right)
-1=\frac{e^{\frac{\Delta_0}{T}}}{2 \cosh^2 \left(\frac{1}{T}\right)}.
\ee
This equation may be written as well in the form:
\bb
\sinh\left(\frac{2}{T}\right) =1+ \frac{1}{2}\eee^{\frac{
\Delta_0}{T}}\,,\;\;
\label{crit2}
\ee
which in turn admits the explicit solution for $\Delta_0$ as function of
$T$ in the form:
\bb
\Delta_{0} =T\,\ln\left[2\sinh\left(\frac{2}{T}\right) -2\right]\,.\;\;
\label{crit3}
\ee
The inverse dependence for $T$ as function of $\Delta_0$ can be evaluated
numerically by solving any of the above equations, which are all equivalent
to the condition of the zero mass in the theory with action
(\ref{SeffCont}). This results the critical line for the BC model shown
in Fig.~1. In the limit $\Delta_0\to-\infty$, from either of the equations
(\ref{criticalline}) and (\ref{crit2}), we recover the Ising case, with
$T_c =2.\,269185$.  For finite $\Delta_0$, as vacancies are added, we
obtain a slowly decreasing (for moderated values of $\Delta_0$) function
for $T_c =T_c(\Delta_0)$, which terminates at the end-point
$(T_c=0,\Delta_0=2)$ at zero temperature, as it can be deduced from
(\ref{criticalline}).  By following the critical line from left to the
right, at first stages, for weak dilution, from physical considerations and
the {\em universality} argument, we expect the transition to be of the
second kind, as in the pure 2D Ising model, while this behavior may be
destructed for sufficiently strong dilution, as $\Delta_0$ increases and
transfer to the positive values, where the correspondent term in the BC
Hamiltonian already suppress the Ising states and is favoring vacancy
states. This happens at a singular point, which we are going to identify
from the condition of stability of the kinetic coefficient in the fermionic
spectrum of the action (\ref{SeffCont}), this is to be discussed in the
next subsection. At the critical line, with zero mass, only derivative
contributions remain in the action Eq. (\ref{SeffCont}). These include the
free fermion kinetic terms, and the ones presenting the residual
interaction between the singlet level and the Ising doublet at the quartic
level in fermions. The HFB decoupling of the interaction will modify the
kinetic part of the action.

%
%
%
%
\begin{figure}[tt!]
\begin{center}
\includegraphics[width=0.8\linewidth]{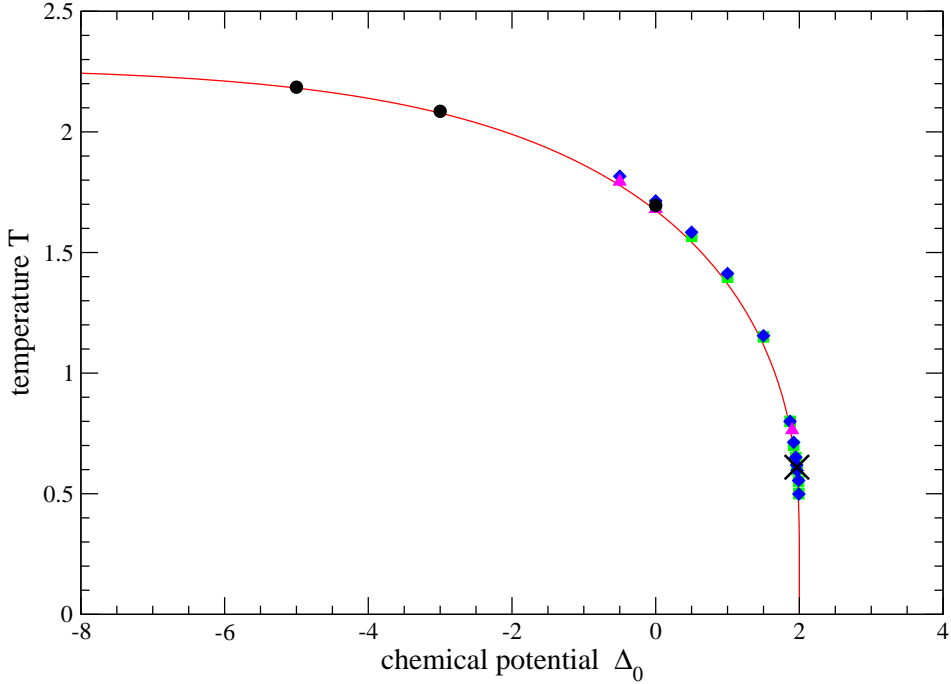}
\unitlength=1cm
\caption{\label{plot} (color online) Comparison between critical line
Eq. (\ref{criticalline}) (plain red line) and numerical results from Monte
Carlo simulations. The black filled dots are from Fig.~1, da Silva
\textit{et al.} \cite{dasilva02} (Wang-Landau method). The cross symbol
indicates the tricritical point identified by the same authors.  The blue
diamond symbols are from Ref. \cite{silva06}, the magenta triangles from
Ref. \cite{xalap98}, and the green squares from Ref. \cite{beale86}
(see also Table 1 for explicit numerical values).}
\end{center}
\end{figure}
%
%

\noindent
The critical line that follows from the condition of zero mass as given
by Eq. (\ref{criticalline}) is plotted in Fig.~\ref{plot} and compared with
recent Monte Carlo simulations by da Silva \textit{et al.} \cite{beale86,
xalap98, liusch02, dasilva02, silva06}. The agreement between numerical
simulations and our results is very good, the mass of the system
(\ref{criticalline}) being exact in that sense, at least in the transition
region. The agreement is within 1\% over the whole range of variation of
$\Delta_0$ at the critical line $T_c =T_c (\Delta_0)$, provided we use the
Monte-Carlo data for $T_c$ as input and evaluate theoretically $\Delta_0$
from (\ref{crit3}) for comparison. The numerical data in the inverse
interpretation, for $T_c=T_c (\Delta_0)$ as a function of $\Delta_0$, are
given in Table 1.  Note that our results are also compatible with exact
upper bound for $T_c(\Delta_0)$ obtained by Braga {\em et al.}
\cite{braga94}. Also notice that the value of $T_c(\Delta_0)$ can be easily
evaluated analytically at the point $\Delta_0=0$, where $\sinh(2/T_c)
=3/2$, with the solution $T_c(0) =1.\,673\,971\,856$. This is to be
compared with the Monte-Carlo results $T_c =1.\,6955\pm 0.0010$, $T_c
=1.\,681(5)$ and $T_c =1.\,714(2)$ \cite{beale86,xalap98,silva06}, the
agreement is again good.

%
%
%
\begin{table}[!tb]
\centering
\begin{tabular}{ c | c c c c }
\hline
$\Delta_0$ &  & Temperature $T_c(\Delta_0)$ &  &
\\
& Ref. \cite{beale86} & Ref. \cite{xalap98} & Ref. \cite{silva06}
(Wang-Landau method)
& Eq. (\ref{criticalline})
\\
\hline
\\
-0.5 & & 1.794(7) & 1.816(2) & 1.7781
\\
0. & 1.695 & 1.681(5) & 1.714(2) & 1.6740
\\
0.5 & 1.567 & & 1.584(1) & 1.5427
\\
1.0 & 1.398 & & 1.413(1) & 1.3695
\\
1.5 & 1.150 & & 1.155(1) & 1.1162
\\
1.87 & 0.800 & & 0.800(3) & 0.7712
\\
1.9 &  & 0.764(7) & 0.755(3) & 0.7221
\\
1.92 & 0.700 & & 0.713(2) & 0.6841
\\
1.95 & 0.650 & & 0.651(2) & 0.6135
\\
1.962 & 0.620 & & 0.619(1) & 0.5776
\\
1.969 & 0.600 & & 0.596(5) & 0.5531
\\
1.99 & 0.550 & & 0.555(2) & 0.4441
\\
1.992 & 0.500 & & 0.499(3) & 0.4270
\\
\hline
\end{tabular}
\caption{
Numerical values of the critical points $(T_c(\Delta_0),\Delta_0)$ in the
BC model: comparison of different numerical simulations and equation
(\ref{criticalline}). Note that small variation of $\Delta_0$ causes more
significant changes in $T_{c}(\Delta_0)$ in the region near $\Delta_0=2$,
as it is to be expected from (\ref{criticalline}).}
\end{table}

%
%
%
\subsection{Tricritical point: Hartree-Fock-Bogoliubov analysis}

The main physical feature of the 2D BC model is the existence of a
tricritical point at the critical line. Below this point, the phase
transition goes from second order to first order: the tricritical point is
characterized by a change in the nature of the singularity. This change
should be seen in the BC spectrum from (\ref{SeffCont}).
In this section, we analyze the effect of the quartic terms in the
action on the stability of the free fermion spectrum at zero mass, along
the critical line $g_0=t^2+2t-1$, by considering the effect of the
interaction part of the action onto the kinetic part within the HFB like
approximating scheme \cite{thouless72,mattuck92,bogoliubov07}. The Ising
part can be easily written in the momentum space representation,
which we will also refer to as Fourier space, after having defined the
following transformations:
\bb
c(\vp)=\frac{1}{L}\sum_{{\vk}}c_{\vk}\exp(i\vk.\vp)\,,\;\;\;\;
\bar c(\vp)=\frac{1}{L}\sum_{{\vk}}\bar c_{\vk}\exp(-i\vk.\vp)\,.\;\;
\label{Four1}
\ee
Using these transformations, the Ising part of the action gains
block-diagonal form,
\bb
\label{SIsingK} \fl
\Sising=\sum_{\vk\in S} it(t+1)(k_x-k_y)(c_{\vk} \bar c_{\vk}
-c_{-\vk} \bar c_{-\vk}) +2itk_xc_{\vk}c_{-\vk}
+2itk_y\bar c_{\vk}\bar c_{-\vk},
\ee
where $S$ is the set of Fourier modes that correspond to half of the
Brillouin zone: if $\vk$ is already included in $S$ then $-\vk$ is not to
be included in $S$ and vice versa (to avoid repetition of modes in the
different sums above), so that couples of modes $(\vk,-\vk)$ fill up the
Brillouin zone exactly once. In fact, terms with $\vk$ and $-\vk$ are
already combined together in (\ref{SIsingK}). The mass term is dropped in
(\ref{SIsingK}) since we are on the critical line. The quartic term can be
written in the Fourier space as
\bb
\Sint=\frac{1}{L^2}\sum_{\vk_1+\vk_2
=\vk_3+\vk_4}V(\vk_2,\vk_4)c_{\vk_1}
c_{\vk_2}\bar c_{\vk_3}\bar c_{\vk_4},
\label{FIint1}
\ee
with the potential
\bb
\nn
V(\vk_2,\vk_4)=-\alpha k_2^xk_4^y+\bet (k_2^xk_4^x+k_2^yk_4^y),\\
\alpha=g_0\,t(t+2\gamma)\,,\;\;\;\; \bet=g_0\,\gamma(1-t)\,.
\label{FPint1}
\ee
Up to now we only expressed the action in the Fourier space, or in the
momentum-space representation, without further approximations. In order to
see if the second order line is stable, we use a mean-field like
approximation  in momentum space, similar to the quantum HFB method.
To do so,
we decompose the fourth order interacting terms into sums of quadratic
terms with coefficients to be determined self-consistently. These
coefficients are actually two-point correlation functions for fermions in
the momentum space. The interaction can be decoupled in different ways.
For example, considering the terms contributing to the Ising action, we may
take account of the averages $\langle c_{\vk} \bar c_{\vk}\rangle $,
$\langle c_{-\vk} \bar c_{-\vk}\rangle $, $\langle c_{\vk}c_{-\vk}\rangle$
and $\langle \bar c_{\vk}\bar c_{-\vk}\rangle $. There are also three
different ways to decouple the interacting term, since $c_{\vk_1}$ can be
paired with either of $c_{\vk_2}$, $\bar c_{\vk_3}$, or $\bar c_{\vk_4}$.
For example,
\bb
c_{\vk_1}c_{\vk_2}=\langle c_{\vk_1}c_{\vk_2}\rangle
+(c_{\vk_1}c_{\vk_2}-\langle c_{\vk_1}c_{\vk_2}\rangle )
\equiv\langle c_{\vk_1}c_{\vk_2}\rangle +\delta_{c_1c_2},
\ee
where $\delta_{c_1c_2}$ is assumed to be a small fluctuation. In this case,
from Eq. (\ref{SIsingK}), the average is non zero only for
$\vk_1 = -\vk_2 =\vk$ or $-\vk$, with $\vk\in S$. We can pair the other
terms by writing the action in the $g_S=3$ different possible ways that are
compatible with the symmetries of Eq. (\ref{SIsingK}), and by using the
fermionic rules, we write:
\bb
\fl
\Sint=\frac{1}{L^2g_S}\sum_{\vk_1+\vk_2=\vk_3+\vk_4}V(\vk_2,\vk_4)
\Big [
(\langle c_{\vk_1}c_{\vk_2}\rangle +\delta_{c_1c_2})(\langle \bar
c_{\vk_3}\bar c_{\vk_4}\rangle +\delta_{\bar c_3\bar c_4})
\;\;\;
\\ \nn
-(\langle c_{\vk_1}\bar c_{\vk_3}\rangle +\delta_{c_1\bar c_3})(\langle
c_{\vk_2}\bar c_{\vk_4}\rangle +\delta_{c_2\bar c_4})
+(\langle c_{\vk_1}\bar c_{\vk_4}\rangle +\delta_{c_1\bar c_4})(\langle
c_{\vk_2}\bar c_{\vk_3}\rangle +\delta_{c_2\bar c_3})
\Big ].
\label{THIS63}
\ee
The next step is to discard terms that are proportional to the squares of
fluctuations $\delta^2$, and keep the others. After some algebra, we obtain
the mean-field quadratic operator for the interaction term as follows:
\bb
\nn \fl
\Sint =\frac{1}{L^2g_S}
\sum_{\vk,\vk' \in S}
4c_{\vk}c_{-\vk}\langle \bar c_{\vk'}\bar c_{-\vk'}\rangle V(\vk,\vk')+
4\bar c_{\vk}\bar c_{-\vk}\langle c_{\vk'}c_{-\vk'}\rangle V(\vk',\vk)
\\ \nn
+c_{\vk}\bar c_{\vk}
\Big(\langle c_{\vk'}\bar c_{\vk'}\rangle v(\vk,\vk')
+\langle c_{-\vk'}\bar c_{-\vk'}\rangle v(\vk,-\vk')
\Big)
\\ \label{MFA}
+c_{-\vk}\bar c_{-\vk}
\Big(
\langle c_{\vk'}\bar c_{\vk'}\rangle v(-\vk,\vk')
+\langle c_{-\vk'}\bar c_{-\vk'}\rangle v(-\vk,-\vk')
\Big),
\ee
where we have defined the potential
\bb
\label{potential}
v(\vk,\vk')=-V(\vk,\vk)-V(\vk',\vk')+V(\vk,\vk')+V(\vk',\vk).
\ee
In the above expressions, there are three different kinds of quantities,
that contribute to the action, associated with the sums like
$\sum_{\vk}c_{\vk} \bar c_{\vk}$, $\sum_{\vk}c_{\vk}\bar c_{\vk}k_i$, or
$\sum_{\vk}c_{\vk} \bar c_{\vk}k_ik_j$, with $i,j=x,y$.  The first term
gives a contribution to the total mass, the second one corresponds to
current operators, and the third one can be thought as a dispersion energy
tensor. Considering the symmetries of the Ising part, and the fact that the
action must be invariant by a dilation factor at criticality, we may only
take into account the current operators. Respectively, we can drop the
first two terms in the potential $v(\vk,\vk')$ defined in Eq.
(\ref{potential}). We define therefore the following unknown parameters,
for the diagonal and nondiagonal couplings of fermions ($i=x,y$):

\bb
\nn
t_i&=&\frac{i}{2L^2}\sum_{\vk\in S}(\langle c_{\vk}\bar c_{\vk}\rangle
-\langle c_{-\vk}\bar c_{-\vk}\rangle )k_i,
\\ \label{parameters}
u_i&=&\frac{i}{L^2}\sum_{\vk\in S}\langle c_{\vk}c_{-\vk}\rangle k_i,
\;\;
\bar u_i=\frac{i}{L^2}\sum_{\vk\in S}\langle \bar c_{\vk}\bar
c_{-\vk}\rangle k_i.
\ee
From the previous discussion, we can drop the first two terms in the
potential $v(\vk,\vk')$ defined in Eq. (\ref{potential}), since we already
assume that only currents are kept as parameters along the critical line.
In this case, it is easy to rewrite, from the property $v(\vk,\vk')
=-v(-\vk,\vk') =-v(\vk,-\vk')$, the effective mean field action of
(\ref{MFA}) as:
\bb
\nn \fl
\Sint=\frac{1}{g_S}
\sum_{\vk\in S}
4ic_{\vk}c_{-\vk}
[(\alpha\bar u_y-\bet\bar u_x)k_x-\bet\bar u_y k_y]
+
4i\bar c_{\vk}\bar c_{-\vk}
[-\bet u_x k_x +(\alpha u_x-\bet u_y)k_y]
\\
+2i(c_{\vk}\bar c_{\vk}-c_{-\vk}\bar c_{-\vk})
[(\alpha t_y-2\bet t_x)k_x+(\alpha t_x-2\bet t_y)k_y].
\;\;\;
\ee
We make then further assumption that, by symmetry invariance in the
momentum space, there exists a solution satisfying $\bar u_y=u_x$,
$\bar u_x=u_y$ and $t_x=-t_y$, so that:
\bb
\nn \fl
\Sint = \frac{1}{g_S}
\sum_{\vk\in S}
4ic_{\vk}c_{-\vk}
[(\alpha u_x-\bet u_y)k_x-\bet u_x k_y]
+
4i\bar c_{\vk}\bar c_{-\vk}
[-\bet u_x k_x+(\alpha u_x-\bet u_y)k_y]
\\
- 2i(c_{\vk}\bar c_{\vk}-c_{-\vk}\bar c_{-\vk})
(k_x-k_y)(\alpha+2\bet)t_x.
\;\;\;
\ee
The total effective action (with zero mass) can finally be written as
\bb
\fl \nn
\action_{\mathrm{eff}}=\sum_{\vk\in S}
i\left[ t(t+1)-\frac{2}{g_S}(\alpha+2\bet)t_x\right](k_x-k_y)
(c_{\vk}\bar c_{\vk}-c_{-\vk}\bar c_{-\vk})
\\ \fl \nn
+
i\frac{4}{g_S}\left[ \left(\frac{g_S}{2}t+(\alpha u_x-\bet
u_y)\right)k_x-\bet u_x k_y\right]c_{\vk}c_{-\vk}
\\ \fl
+
i\frac{4}{g_S}\left[-\bet u_x k_x+\left(\frac{g_S}{2}t+(\alpha u_x-\bet
u_y)\right)k_y\right]\bar c_{\vk}\bar c_{-\vk},
\ee
or in a more compact form as
\bb
\fl\nn
\action_{\mathrm{eff}}
=\sum_{\vk\in S}
ic(k_x-k_y)(c_{\vk}\bar c_{\vk}-c_{-\vk}\bar c_{-\vk})
+
2i(ak_x-bk_y)c_{\vk}c_{-\vk}
\\
+
2i(-bk_x+ak_y)\bar c_{\vk}\bar c_{-\vk},
\label{SEFF70}
\ee
with the following coefficients
\bb\fl
a=t+2\frac{\alpha u_x-\bet u_y}{g_S}, \;\;\;\;\;
b=2\bet \frac{u_x}{g_S}, \;\;\;\;\;
c=t(t+1)-2t_x\frac{\alpha+2\bet}{g_S}.
\ee
The partition function can then be written as a product over the Fourier
modes $Z=\prod_{\vk\in S}Z_{\vk}$, with
\bb
\label{fpFourier}
Z_{\vk}=k^2[A+B\sin 2\theta_k],
\ee
$\theta_k$ being the angle of the vector $\vk$, and
\bb
A=c^2-4ab,\:\:
B=-c^2+2(a^2+b^2).
\ee
We assume that $|A|$ is larger than $|B|$ on the second order critical
line, until a singular point is reached, where eventually $A^2=B^2$.
Indeed, the expression (\ref{fpFourier}) is valid only if the elements
$A+B\sin 2\theta_k$ are all strictly positive, which is the case only if
$A^2>B^2$.  This will be checked using numerical analysis. Beyond this
point, the effective action is unstable and has to be modified to
incorporate further corrections. In a bosonic $\Phi^6$ Ginzburg-Landau
theory describing a first order transition, the tricritical point is
usually defined as the point where both coefficients of $\Phi^2$ and
$\Phi^4$ terms vanish \cite{lawrie84,zj04}. By analogy, in the present
fermionic theory, it is tempting to associate the above singular
point with the effective tricritical point.\\
The parameters $t_x$, $u_x$ and $u_y$ are to be determined
self-consistently from the definitions Eqs. (\ref{parameters}). In the
continuous limit, these reduce to

\begin{figure}
\begin{center}
\includegraphics[width=0.8\linewidth]{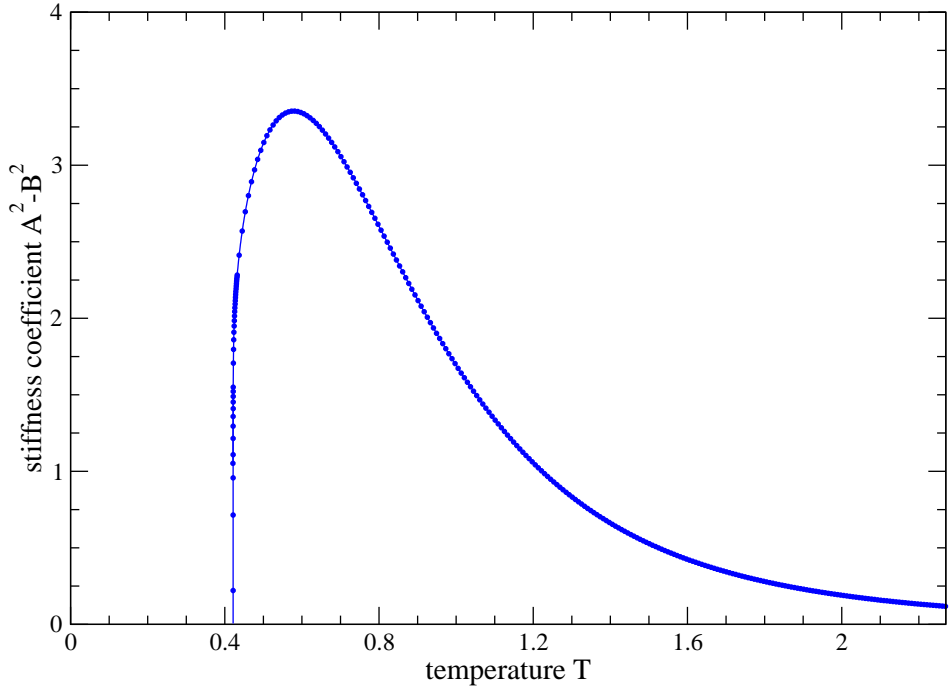}
\unitlength=1cm
\caption{\label{tric1} (color online) Stiffness of the spectrum: solution
of HFB self-consistent Eqs. (\ref{relations}) for the
coefficient $A(T)^2-B(T)^2$ as function of $T$. The temperature where
$A(T)^2-B(T)^2=0$ gives the location of the singular
point $T_{\mathrm{t}}^{*}$.}
\end{center}
\end{figure}

\bb
\nn
t_x&=&\frac{c}{4\pi}\int_0^{\pi}\dd \theta\frac{1-\sin 2\theta}{A+B\sin
2\theta},
\\
u_x&=&\frac{1}{2\pi}\int_0^{\pi}\dd \theta\frac{a\sin 2\theta-b}{A+B\sin
2\theta},
\;\;
u_y=\frac{1}{2\pi}\int_0^{\pi}\dd \theta\frac{a-b\sin 2\theta}{A+B\sin
2\theta}.
\ee
After computing the trigonometric integrals, we obtain the relations
\bb\nn
t_x=\frac{c}{4B}\left(
-1+(A+B)\frac{{\rm sign}(A)}{\sqrt{A^2-B^2}}
\right),
\\ \nn
u_x=\frac{1}{2B}\left(
a-(aA+bB)\frac{{\rm sign}(A)}{\sqrt{A^2-B^2}}
\right),
\\ \label{relations}
u_y=\frac{1}{2B}\left(
-b+(bA+aB)\frac{{\rm sign}(A)}{\sqrt{A^2-B^2}}
\right).
\ee
Numerically, we proceed the following way. Starting from
$T$ slightly below $T_c(-\infty)$, we solve the consistency equations for
$t_x$, $u_x$ and $u_y$, with the value of $\Delta_0$ given by the critical
line (\ref{criticalline}) at a given temperature. The solutions are then
plug into the coefficients $A(T)$ and $B(T)$, and we plot $A(T)^2-B(T)^2$
as a function of $T$, as is shown in figure~\ref{tric1}. We repeat the
process by decreasing the temperature until we reach the point where this
quantity vanishes.\\
By doing so we find a singular point approximately located at
$(T_{\mathrm{t}}^{*},\Delta_{0,\mathrm{t}}^{*})\simeq(0.42158, 1.9926)$.
This is close to the tricritical point $T_{\mathrm{t}}$ given by Monte
Carlo simulations:  $(T_{\mathrm{t}},\Delta_{0,\mathrm{t}})\simeq(0.610,
1.9655)$ \cite{dasilva02}, and $(T_\mathrm{t},\Delta_{0,\mathrm{t}})
\simeq(0.609(3), 1.966(2))$ \cite{silva06}. If we assume that
$T_{\mathrm{t}}^{*}$ represents the tricritical point, the mean-field like
treatment of the underlying field theory underestimates the fluctuations,
rendering the second order critical line more stable at lower temperatures,
as compared to Monte-Carlo results, as we approach $(T_c=0,\Delta_0=2)$
along the critical line. Stronger fluctuations can be simulated by
lowering the value of $g_S$, which increases (lowers) the value of
$T_{\mathrm{t}}^{*}$ ($\Delta_{0,\mathrm{t}}^{*}$), respectively. Instead
of $g_S=3$, taking $g_S=2.5$, for example, leads to a
$T_{\mathrm{t}}^{*}\simeq 0.48$, closer to the Monte Carlo results. This
can be achieved precisely by incorporating more diagrams in the computation
of the effective free energy \cite{mattuck92}. Also, due to the fact that
we are in a region near $(T_c=0,\Delta_0=2)$, where the change in
temperature is large compared to the change of $\Delta_0$ (the slope is
vertical at this point as is seen in figure~\ref{plot}), it is more
difficult to obtain a precise value of $T_{\mathrm{t}}^{*}$ within a
mean-field treatment.\\
It is important that the BC fermionic action (\ref{SeffCont}) finally
predicts the existence of a special (tricritical) point at the critical
line somewhere close (in $\Delta_0$) to termination point of that line at
$(T_c=0,\Delta_0=2)$. The tricritical point is defined, within this
interpretation, as the point of the destruction, or loss of stability, in
the effective fermionic spectrum of the action due to the modifications
introduced into the kinetic part by a sufficiently strong dilution of a
system by the vacancy states, which corresponds to large enough coupling
constant $g_0$, as it was commented above.
\footnote{ \
It may be also noted that the Monte-Carlo values for
$(T_\mathrm{t}, \Delta_{0,\mathrm{t}})$ seemingly lie practically on the
theoretical curve for the critical line (\ref{criticalline})-(\ref{crit3}).
For instance, taking as input value $T_\mathrm{t}\simeq 0.\,609(3)$
\cite{silva06}, from (\ref{crit3}) we find $\Delta_{0,\mathrm{t}}\simeq
1.952$, which is sufficiently close to the M-C value $\Delta_{0,\mathrm{t}}
\simeq 1.\,966(2)$ from this set \cite{silva06}, the deviation being
probably less than 1\%.}.

%
%
%
%
%

%
%
\section{Conclusions}

In this paper, we have considered the physics of the BC model as a
fermionic field theory.  Using Grassmann algebra, we have shown that the
model can be transformed into quantum field theoretical language in terms
of fermions alias Grassmann variables. This fermionic theory for BC model
is described by an exact fermionic action with an interaction on a discrete
lattice. This action can be reduced, after some transformations, in the
continuum limit and low energy sector, to an effective continuum field
theory which includes a modified Ising action, which is quadratic in
fermions, and a quartic interaction. From there we have extracted the exact
mass of the model and analyzed the effect of the quartic term on the
stability of the free fermion spectrum in the kinetic part. The condition
of the zero BC mass gives the critical line of phase transition
points in the $(T,\Delta_0)$ plane, which is found to be in a very good
agreement with the results of Monte-Carlo simulations over the whole range
of variation of concentration of the non-magnetic sites governed by
$\Delta_0$. The location of the tricritical point needs additional analysis
of the excitation spectrum of integral factors $Z_{\vk}$ of $Z$ around the
origin in the momentum space. In particular, the \textit{stiffness} of the
excitation spectrum (the coefficient in front of $\vk^2$ term in
factors $Z_{\vk}$ as we expand the dispersion relation for $Z$ in momentum
variables) vanishes at a singular point $T_{\mathrm{t}}^{*}$, which we
assume to be identified as the tricritical point $T_{\mathrm t}$. A
Hartree-Fock-Bogoliubov analysis gives an approximate location for this
point on the phase diagram (critical line) which can be compared to
the numerical results of Monte Carlo simulations. The more precise location
of the instability point could be achieved by taking into account more
diagrams contributing to the effective free energy. In any case, we have
shown the existence of a singular point at the critical line by studying
the stability of the kinetic spectrum of the action at this line, where the
nature of the transition is to be changed due to strong dilution. The main
result of this paper is the possibility to study precisely first-order
transition driven systems from a fermionic point of view using Grassmann
algebra. The method we have applied may be useful as well for other systems
where effective field theory is presented by an action similar to that of
Eq. (\ref{SeffCont}).  In essence, this is a one of the simplest form of an
action with 4-fermion interaction that can be written out from a unique
pair of Grassmann variables at each point of the real space in two
dimensions. Application of the same method to other extensions of the BC
Hamiltonian, such as the Blume-Emery-Griffiths model \cite{BeG71}, is also
possible. Finally, at intermediate stages, a partial bosonization of the
system leads to a $mixed$ representation of the model not only in term of
fermions but also in term of hard core {\em bosons}, as written explicitly
in the lattice action of Eq. (\ref{act1fin}). The representations of this
kind could be useful also to look for a possible interpretation of the
tricritical point in the BC model as a special point in the phase diagram
where an additional hidden symmetry between fermions and bosons may appear.

\end{document}